\documentclass[12pt,letterpaper]{article}
\usepackage{epsfig,rotating,setspace,latexsym,amsmath,epsf,amssymb,amsfonts,bm,theorem,cite,caption,subcaption,enumerate,longtable,accents,algorithm,graphicx,epsf,authblk,epstopdf,url,color,multirow,algpseudocode,mathtools,comment,tabularx,comment,physics,graphicx,caption,subcaption,braket,enumitem,xcolor,algorithm, dsfont}
\usepackage[short,c2]{optidef}
\usepackage[toc,title]{appendix}
\usepackage{mathrsfs}

\newtheorem{theorem}{Theorem}

\newtheorem{corollary}{Corollary}
\newtheorem{definition}{Definition}
\newtheorem{remark}{Remark}
\newtheorem{lemma}{Lemma}
\newtheorem{example}{Example}

\newenvironment{Proof}[1]{\medskip\par\noindent{\bf Proof:\,}\,#1}{{\mbox{\,$\blacksquare$}\par}}

\allowdisplaybreaks

\title{Local Private Information Retrieval for Graph-Based Replicated Systems\footnote{A part of this work was accepted in IEEE ITW 2026.}}

\author{Shreya Meel \qquad Mohamed Nomeir \qquad Sennur Ulukus \\
\normalsize Department of Electrical and Computer Engineering \\
\normalsize University of Maryland, College Park, MD 20742 \\
\normalsize {\qquad {\it smeel@umd.edu} \qquad \it mnomeir@umd.edu} \qquad {\it ulukus@umd.edu}}

\begin{document}
\date{}
\maketitle
\vspace{-0.8cm}

\begin{abstract}
    We rethink the definition of privacy in multi-server, graph-replicated private information retrieval (PIR) systems, by introducing a novel setting where the user's privacy is governed by the servers' storage structure. In classical graph-replicated PIR, the user retrieves a single message stored at the servers, while hiding the message index from each individual server. In our proposed privacy setting, the user is concerned with hiding the message index from a particular server, only if that server stores the message being retrieved, and privacy is not imposed otherwise. We coin this relaxed privacy requirement as \emph{local user privacy} and the resulting PIR problem as \emph{local PIR} on the graph. Our focus is on two-replicated PIR systems, where every message is replicated twice and stored on two distinct servers. Specifically, we study local PIR systems where the storage is represented by simple graphs, i.e., every pair of vertices is associated with at most one edge, and by their multigraph extension, i.e., $r$ parallel edges replace every edge. For these settings, we establish bounds on the local PIR capacity, defined as the maximum number of message symbols retrieved, per downloaded symbol. The local privacy requirement yields significant capacity gain over the classical PIR capacity under the same storage structure. For instance, in settings where the graph is a disjoint union of multiple identical sub-graphs, the gain in the local PIR capacity over classical PIR capacity is multiplicative in the number of sub-graphs. Further, for connected graphs, we derive capacity lower bounds for edge-transitive and bipartite graphs, which are greater than the best-known PIR capacity bounds. From these and by establishing matching upper bounds, we exactly characterize the capacity for star graphs, cyclic graphs, and path graphs with odd number of vertices. We also introduce two local PIR schemes, one of which supports single-symbol messages, for general simple graphs. Finally, we extend these results to the case of multigraphs, and propose two achievable schemes for general $r$-multigraphs, along with explicit upper-bounds for cyclic and path multigraphs.
\end{abstract}

\newpage\section{Introduction}
Private information retrieval (PIR) studies the fundamental problem of efficiently retrieving a desired message $W_{\theta}$ from an indexed database, typically replicated across multiple, non-colluding servers, while ensuring that the identity of $\theta$, i.e., the message index, is concealed from each individual server. This problem is motivated by crucial applications, such as those in medicine, finance, national defense, where revealing the user's requested index reveals sensitive information such as the user's interest or intent. The PIR system model consists of $N$ servers storing $K$ messages, whose contents the user is unaware of. For retrieving $W_{\theta}$, the user communicates with the servers by generating and sending queries, to which the servers respond by constructing answers. From these answers, the user can correctly retrieve $W_{\theta}$, while each server remains oblivious as to which of the $K$ messages the user has retrieved. The PIR problem was introduced in the seminal paper \cite{chor}, and was further studied in the computer science community (see e.g.,\cite{beimel_pir,  BIKR02, woodruff2005geometric, beimel2007robust, dvirgopi_2server}) with the focus on optimizing the communication and computation costs incurred, where each message is treated as a single bit. Thereafter, PIR garnered renewed interest in the information theory community, where the message lengths are assumed to be arbitrarily large, and the goal is to minimize the amount of information that the user downloads from the servers. Consequently, the capacity of PIR, defined as the maximum attainable ratio between the message size and the amount of downloaded information, became the metric of interest. The PIR capacity was first derived in \cite{SJ17} for the fully-replicated database setting, where each of the $N$ non-colluding servers stores all $K$ messages, after which another capacity-achieving scheme, which also guarantees minimal subpacketization, was proposed in \cite{ChaoTian}. This was extended to more realistic scenarios, such as server-dropouts \cite{rashmi_erasure_pir, Salim_CodedPIR , mdstpir, coded_colluding_2017, banawan_pir_mdscoded, tspir_mdscoded}, to address further security aspects, such as database security \cite{c_spir,  skoglund_mds_spir, zhusheng_spir_pir}, server collusion \cite{colluding, pir_spir_adversaries, arbitrarycollusion,  csa, nomeirasymmetric, semantic_tpir}, eavesdroppers \cite{sun_eaves, nan_eaves, C_SETPIR}, malicious servers \cite{byzantine_tpir,  nomeir_asymp_bspir},  and recently to accommodate quantum communication channels between the servers and the user \cite{hayashi_qpir,  hollanti_qpir,  jafar_qpir,  ulukus_qpir}.

However, as the amount of data keeps growing at a rapid pace, the assumption that all messages are stored at all servers is becoming expensive and unrealistic. Moreover, geographical and access constraints may restrict the storage of some data at certain servers. Single-server PIR \cite{kadhe_singleserver_pir, single-serverSI} is one approach to solve this, but it requires availability of side information at the user, to accomplish a PIR scheme other than downloading all messages. To circumvent this requirement, the study of PIR with non-fully-replicated databases was initiated in \cite{graphbased_pir,  BU19, asymp_gxstpir}, where the storage of messages at the servers is modeled by an undirected hyper-graph. Specifically, the servers are represented by vertices, and the messages are represented by hyper-edges. A copy of a message is stored at servers if and only if the hyper-edge representing the message is incident to the vertices representing the corresponding servers. To obtain more tangible results, this model was specialized to two-replicated systems, where each edge is incident to exactly two vertices, leading to exact and approximate PIR capacity results for common families of simple graphs \cite{SGT23, YJ23,  our_journal2025,  krishnan_graph, gePIR}  and $r$-multigraphs \cite{meel_multi_pir,  gePIR, our_journal2025, krishnan_graph} where each edge is replaced by $r$ parallel edges. Recently, in \cite{meel2025symmetric,  shreya_common_randomness_gspir, meelrolecommonrandomnessreplication}, these results were extended to address database security, under graph-replicated and fully-replicated common randomness settings.

In all the hypergraph-replicated PIR models, the PIR capacity decreases as the number of servers $N$ (i.e., the number of vertices in a given hypergraph) increases, contrary to the fully-replicated case. Apart from partial replication of messages, this is also because, the privacy requirement has been kept the same as the one in the fully-replicated formulation in \cite{SJ17}. In other words, the usual model mandates the retrieval of all messages to be uniformly private against all servers, for all message indices, irrespective of whether the given message is stored at that server or not. This may be too restrictive towards the goal of preventing user-profiling, since the said server is unaware of the message contents. From a practical standpoint, if a server does not store a message, knowing that the user has requested that message is unlikely to breach the user's privacy. From this viewpoint, we introduce a new perspective to graph-replicated PIR, by studying \emph{local PIR}, wherein the only message indices that are kept private from a given server are those that are stored in it. This relaxes the privacy constraint, and increases the capacity for a given graph. Note that, our formulation extends seamlessly to the fully-replicated setting, where local PIR becomes equivalent to the original PIR formulation. 

Related to our work is the PIR-variant that relaxes perfect privacy to an acceptable privacy budget. In this vein, leaky PIR\cite{tandon_leaky} and weakly private PIR \cite{tian_leakage_pir, improved_weakly, multi_server_weak, weak_pir_graphs} have been studied. The goal there is to maximize the PIR rate, under a differential privacy, maximal leakage or mutual information leakage budget. In contrast, our proposed privacy relaxation arises naturally from the storage architecture of non-fully-replicated systems, while maintaining perfect privacy for the subset of indices that are stored in each server. In other words, in a given instance of local PIR, we expect perfect privacy for a desired message index against the servers that store that message. From the remaining servers, we seek to directly download a minimum amount of side information, so that the PIR rate is maximized.

In this work, we first focus on deriving local PIR capacity bounds for graphs whose edges are incident to only two vertices, and compare them with the standard graph-based PIR results in the literature. Our contributions for simple graphs are as follows. Our first prominent result is the local PIR capacity of graphs that are (vertex-)disjoint union of $m$ connected subgraphs. If the subgraphs are identical, the local PIR capacity improves upon that of standard PIR by at least a factor of $m$. Next, we focus on connected graphs, i.e., those graphs which have a path between every vertex pair. For edge-transitive graphs, we derive capacity lower bounds for PIR systems based on these graph families, such as cyclic, complete, and star graphs. Using this result, and a matching converse, we show that the local PIR capacity for the cyclic graph with $N$ vertices is $\frac{1}{2}$, i.e., independent of the number of servers, while its PIR capacity is $\frac{2}{N+1}$\cite{BU19}. Starting from the scheme of complete graphs, we derive a local PIR scheme construction that yields a general capacity lower bound for any simple, connected graph. We follow this up with another scheme for general graphs, which is not necessarily better than the previous scheme in terms of rate, but offers more flexibility by requiring minimal subpacketization. Under the local privacy requirement, we derive a simple yet effective scheme for the class of bipartite graphs. This scheme achieves the local PIR capacity of $1$ for star graphs, which is superior to the PIR capacity by a factor of $\sqrt{N}$ for large $N$. Using the scheme, and a matching converse, the local PIR capacity for path graphs is derived as $\frac{N-1}{2N-4}$ for odd $N$, while its PIR capacity is $\frac{2}{N}$\cite{our_journal2025}. Finally, we generalize our results to $r$-multigraphs, that extend the two proposed schemes on simple graphs. Interestingly, both schemes achieve their respective simple graph local PIR rates multiplied by $(2-2^{1-r})^{-1}$. However, one of them incurs expansion of subpacketization by a factor of $2^{r-1}$, while the other maintains the subpacketization of $1$. 

The rest of the paper is organized as follows. In Section~\ref{sec:sys_mod}, we present the new problem formulation for the local PIR problem for simple graphs, along with examples and graph-theoretic definitions. In Section~\ref{sec:simple graph results}, we state the main results for simple graphs and their consequences. In Section~\ref{sec:ach proofs}, we develop the schemes that yield the capacity lower bounds, while in Section~\ref{sec:ubnd proofs}, the corresponding upper bounds are derived. In Section~\ref{sec:multigraphs}, we extend the results to the multigraph version of simple graphs. Finally, Section~\ref{sec:conclude} concludes the paper.

\textbf{Notations:} To denote the set of integers $\{1,2,\ldots,n\}$ for a positive integer $n$, we use the notation $[n]$. For positive integers $m,n$ with $m\leq n$, the notation $[m:n]$ represents the set $\{m,m+1,\ldots,n-1,n\}$. We assume every vector $\bm{x}$ to be a column vector, where $\bm{x}^\top$ is its transposed row vector. For a matrix $\bm{X}$, the vectors $\bm{X}(i,:)$ and $\bm{X}(:,j)$ denote its $i$th row and $j$th column, respectively. For a set or family of sets $\mathcal{X}$, we use $|\cdot|$ to denote its cardinality.

\section{System Model and Preliminaries}\label{sec:sys_mod}
In this section, we start by describing the PIR system model for simple graphs in an information-theoretic framework, followed by motivating examples to introduce the concept of local privacy. Then, we provide important graph-theoretic definitions that are required in the main results section.

\subsection{Problem Setting: Simple Graphs}
In the traditional fully-replicated PIR setting, each message $W_k$, $k \in [K]$, is replicated in all servers, i.e., if $\mathcal{W}_n$ is the storage at the $n$th server, then $\mathcal{W}_n = \{W_1, W_2, \ldots, W_K\}$, $n\in [N]$. On the other hand, in graph-based/graph-replicated PIR, each server stores only a subset of the messages, i.e., $\mathcal{W}_n \subseteq \{W_1, W_2, \ldots, W_K\}$ for server $n$. In both cases, it is assumed that each message is a vector of $L$ independent symbols generated uniformly at random from the field $\mathbb{F}_q$, where $q$ is a prime power, and that the messages are independent of each other. The joint entropy of all messages in the PIR system is
\begin{align}
    H(W_{1}, W_2, \ldots, W_K) = \sum_{i=1}^K H(W_i) = KL, \quad \text{in $q$-ary units.}
\end{align}
Given a simple, undirected graph $G = (V,E)$, where $V$ denotes the set of nodes/vertices and $E$ denotes the set of edges, the respective PIR database system consists of $N=|V|$ servers and $K=|E|$ messages. Each message is replicated exactly twice and stored on two distinct servers in $[N]$, such that an edge incident to a particular vertex means that the message represented by the edge is stored at the respective server. Specifically, let the message $W_k$ be stored at servers $i$ and $j$. Then, the edge $\{i,j\}\in E$ incident with servers $i$ and $j$, is assigned a unique index $k\in[K]$. Let us denote the set of message indices in the $n$th server as
\begin{align}
    \mathcal{I}_n = \{\ell \in [K]: W_{\ell} \in \mathcal{W}_n\} \subseteq [K].
\end{align}
Then, each server stores exactly $\deg(n)$ messages, where $\deg(n)=|\mathcal{I}_n|$ is the degree of node $n$ in $G$. 

Let $\theta \in [K]$ be the message index required by the user and $Q_n^{[\theta]}$ be the query sent by the user to the $n$th server to retrieve $W_{\theta}$. Let $\mathcal{Q}$ denote all the queries sent by the user to all servers across all message indices, i.e., $\mathcal{Q} = \{Q_n^{[\theta]}: n\in [N], ~ \theta \in [K]\}$. Then, $Q_n^{[\theta]}$, being a part of $\mathcal{Q}$ can be determined exactly given $\mathcal{Q}$. Since the user has no prior knowledge about the message content, the queries are independent from them, i.e., 
\begin{align}\label{r_1}
    I(\mathcal{Q}; W_1, \ldots, W_K) = 0.
\end{align}
In the typical graph-based PIR studied so far in the literature, where the required message index has to be private from all the servers, i.e., 
\begin{align}\label{r_pir}
    I(\theta;Q_n^{[\theta]}) = 0, \quad n \in [N].
\end{align}
However, in local PIR, we focus on the local privacy requirement only, making the privacy notion less restrictive. More specifically, we only keep the index private from the servers storing the required message, i.e.,
\begin{align}\label{r_2}
    I(\theta; Q_n^{[\theta]}~| ~\theta \in \mathcal{I}_n) = 0, \quad n\in [N].
\end{align}
This relaxation of privacy is the main difference between classical graph-based PIR and local PIR introduced here. As in typical PIR, we assume that the servers are honest but curious, i.e., if the $n$th server receives a query $Q_n^{[\theta]}$, it generates an answer $A_n^{[\theta]}$, as a deterministic function of the received query and the stored message symbols, which implies,
\begin{align}\label{r_3}
    H(A_n^{[\theta]}|\mathcal{W}_n,Q_n^{[\theta]} ) = 0.
\end{align}
Once the user receives the answers from all the servers,\footnote{We assume without loss of generality that when a query is not sent to a server, the set of answers from this server is $\emptyset$.} the required message must be exactly decodable without errors, i.e.,
\begin{align}
    \label{r_4}H(W_{\theta}|A_{[N]}^{[\theta]}, \mathcal{Q}) = 0. 
\end{align}
A \emph{local PIR} scheme $\Pi$ on a graph $G$ is identified by the set of queries $\mathcal{Q}$, and the corresponding set of answers $\mathcal{A} = \{A_n^{[\theta]}, n\in [N], \theta \in [K]\}$ which satisfy \eqref{r_1}, \eqref{r_2}, \eqref{r_3} and \eqref{r_4}. 

As in the PIR literature, $L$ is referred to as the subpacketization of the scheme $\Pi$. Let the number of downloaded symbols to retrieve the $k$th message $W_k$ be denoted as $D_k$, i.e.,
\begin{align}
    D_k = \sum_{n=1}^N H(A_n^{[k]}).
\end{align}
Let $\bm{P}=(\mathbb{P}\left(\theta = k\right), ~k\in [K])$ be the probability mass function of the desired message index. Then, the rate of the scheme $\Pi$ for a graph $G$ is defined as 
\begin{align}
    R^{\Pi}_{\bm{P}}(G) = \frac{L}{\mathbb{E}[D]}=\frac{L}{\sum_{k=1}^K \mathbb{P}(\theta=k)D_k}.
\end{align}
Throughout the paper, we assume that $\theta$ is uniform over $[K]$, simplifying the local PIR rate expression to
\begin{align}\label{eq:rate_uniform}
    R^{\Pi}(G) =\frac{KL}{\sum_{k=1}^K D_k}.
\end{align}
The capacity of the local PIR for graph $G$ is given by the supremum of rates \eqref{eq:rate_uniform} over all achievable local PIR schemes $\Pi$, i.e.,
\begin{align}
    C(G) = \sup_{\Pi} \ R^{\Pi}(G).
\end{align}
In the following, we write $C(G)$ as the local PIR capacity for the graph $G$, and $C_{PIR}(G)$ as the PIR capacity for the same graph.

\begin{remark}
    Due to the relaxed privacy constraint in \eqref{r_2} with respect to \eqref{r_pir}, the inequality $C(G)\geq C_{PIR}(G)$, where $C_{PIR}$ holds for any graph $G$.  
\end{remark}

\subsection{Motivating Examples}\label{motive_ex}
We provide the following examples to present the main ideas of the proposed local PIR formulation, while considering storage settings based on common simple graph families.

\begin{figure}[t]
    \centering
    \subfloat[$\mathbf{C}_4$]
    {\includegraphics[width=0.22\textwidth]{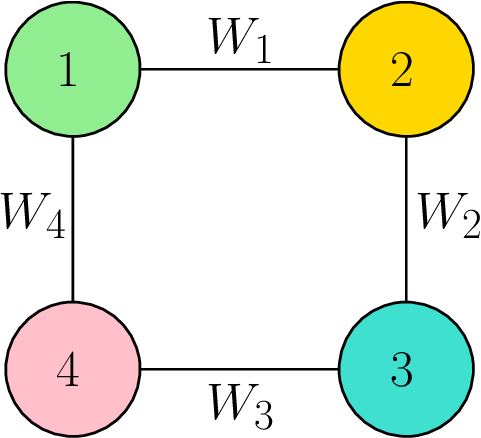}
    \label{fig:C_4}}
    \hspace{1cm}
    \subfloat[$\mathbf{K}_4$]
    {\includegraphics[width=0.22\textwidth]{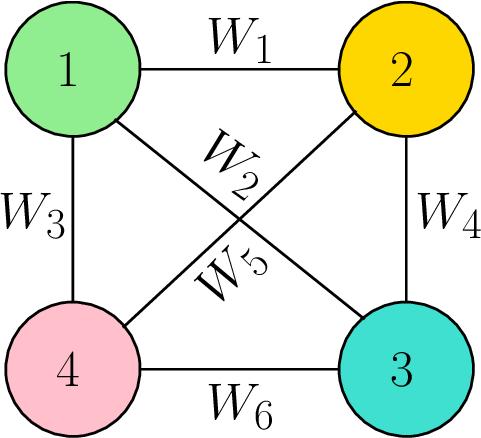}
    \label{fig:K_4}}
     \hspace{1cm}
    \subfloat[$\mathbf{K}_{2,3}$]
    {\includegraphics[width=0.22\textwidth]{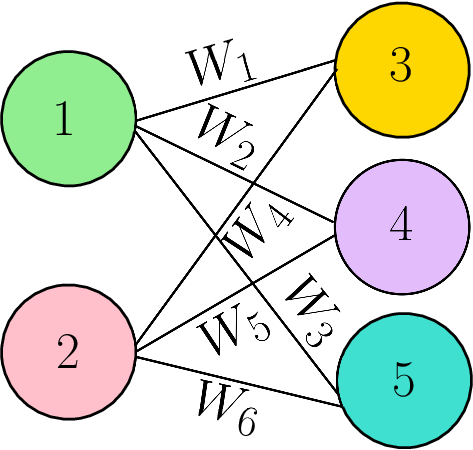}
    \label{fig:K_2,3}}
    \label{example_graphs}
    \caption{Cyclic, complete and complete bipartite graphs. The vertices represent servers, and the edges represent the messages replicated at these servers.}
\end{figure}

\begin{example}[Cyclic Graph]\label{ex_cycle}
Consider a cyclic graph $\mathbf{C}_N$ for $N=4$ servers (see Fig.~\ref{fig:C_4}) with the storage in Table \ref{c_4_storage_table}, i.e., $\mathcal{W}_1 = \{W_1,W_4\}$, $\mathcal{W}_2 = \{W_1,W_2\}$, $\mathcal{W}_3 = \{W_2,W_3\}$, $\mathcal{W}_4 = \{W_3,W_4\}$.

\begin{table}[h]
    \centering
    \setlength{\tabcolsep}{5pt}
    \begin{tabular}{|c|c|c|c|}
    \hline
    server 1& server 2& server 3& server 4\\
    \hline
      $W_4$& $W_1$ & $W_2$ & $W_3$\\
     $W_1$ & $W_2$ & $W_3$ & $W_4$\\
        \hline
    \end{tabular}
    \caption{Storage for each server for $\mathbf{C}_4$ in Example~\ref{ex_cycle}.}
    \label{c_4_storage_table}
\end{table}

Let the number of symbols per message (subpacketization) be $L=2$. Thus, each message can be represented as $W_1=(a_1,a_2)$, $W_2=(b_1,b_2)$, $W_3 = (c_1,c_2)$ and $W_4=(d_1,d_2)$ after the user permutes the message symbols independently and uniformly at random. 

Table~\ref{tab:C4 answers} illustrates the scheme for local PIR for each required message index. We make a couple of important observations, focusing on a single server, say server $1$: First, we note that for messages stored in server 1, i.e., $\theta=1$ and $\theta=4$, the queries are identical, providing privacy. Second, we note that even when the desired message is not stored in this server, i.e., $\theta=2$ and $\theta=3$, the user still queries the server to download side information to be used together with downloaded information from other databases. 

\begin{table}[h]
\centering
    \begin{tabular}{|c|c|c|c|c|}
    \hline
        & server 1& server 2& server 3& server 4\\
        \hline
        $\theta = 1$& $a_1+d_1$ & $a_2+b_1$ & $b_1$ & $d_1$\\
        $\theta = 2$& $a_1$ & $a_1+b_1$ & $b_2+c_1$ & $c_1$\\
        $\theta = 3$& $d_1$ & $b_1$ & $b_1+c_1$ & $c_2+d_1$\\
        $\theta = 4$& $a_1+d_1$ & $a_1$ & $c_1$ & $c_1+d_2$\\
        \hline
    \end{tabular}
    \caption{Retrieval scheme for $\mathbf{C}_4$ in Example~\ref{ex_cycle}.}
    \label{tab:C4 answers}
\end{table}

The scheme downloads one symbol from each database, thus its rate is $\frac{2}{4} =\frac{1}{2}$. In contrast, the PIR capacity for the same setting \cite{BU19} is given by $C_{PIR}(\mathbf{C}_4) = \frac{2}{5} < \frac{1}{2} = C(\mathbf{C}_4)$.
\end{example}

\begin{example}[Star Graph]\label{exmp:star_graph}
In the star graph $\mathbf{S}_N$, we have a central node (say server $N$) storing all the messages and each of the remaining servers stores only one of those messages. That is, $K=N-1$, and server $n\in [N-1]$ stores the message $W_n$. 

The local PIR scheme for $\mathbf{S}_N$ is given as follows. If $\theta = k$, the user contacts only the $k$th server, and downloads the entire message $W_k$. Since server $N$ is never contacted to download a message, it stays oblivious of the desired message index $\theta$. This yields the rate as $1$, whereas, the PIR capacity for star graphs is $C_{PIR}(\mathbf{S}_N) = \Theta\left(\frac{1}{\sqrt{N}}\right)$ \cite{SGT23}.
\end{example}

\begin{example}[Complete Graph]\label{ex_complete}
For the complete graph with $N$ vertices $\mathbf{K}_N$, each node shares an edge with all other vertices, i.e., the degree of each node is $N-1$ and the total number of edges is $K=\binom{N}{2}$. We consider the case of $N=4$ servers, as depicted in Fig.~\ref{fig:K_4}, with the storage given in Table~\ref{k_4_storage}.

\begin{table}[h]
    \centering
    \begin{tabular}{|c|c|c|c|}
    \hline
          server 1& server 2 & server 3 & server 4\\
         \hline
            $W_1$& $W_1$ & $W_2$ & $W_3$\\
            $W_2$& $W_4$ & $W_4$ & $W_5$\\
            $W_3$& $W_5$ & $W_6$ & $W_6$\\
         \hline
    \end{tabular}
    \caption{Storage for $\mathbf{K}_4$ in Example~\ref{ex_complete}.}
    \label{k_4_storage}
\end{table}

Let the number of symbols per message be $L=4$. Thus, the permuted message symbols can be represented as $W_1=(a_1,\ldots,a_4)$, $W_2=(b_1,\ldots,b_4)$, $W_3=(c_1,\ldots,c_4)$, $W_4 = (d_1, \ldots, d_4)$, $W_5 = (e_1, \ldots,e_4)$ and $W_6 = (f_1, \ldots, f_4)$.

\begin{table}[h]
\centering
    \begin{tabular}{|c|c|c|c|c|}
    \hline
    & server 1 & server 2 & server 3& server 4\\
    \hline
    \multirow{3}{*}{$\theta =1$} & $a_1+b_1$ & $a_3+d_1$ & $b_1$ & $c_1$\\
    & $a_2+c_1$  &  $a_4+e_1$& $d_1$ & $e_1$\\
    & $b_2 +c_2$ & $d_2+e_2$ & &\\
    \hline
    \multirow{3}{*}{$\theta =2$} & $a_1+b_1$ & $a_1$ & $b_3+d_1$ & $c_2$ \\
    & $a_2+c_1$ & $e_1$ & $b_4+f_1$ & $f_1$\\
    & $b_2+c_2$ & & $d_2+f_2$ &\\
    \hline
     \multirow{3}{*}{$\theta =3$} & $a_1+b_1$ & $a_2$ & $b_2$ & $c_3+e_1$ \\
    & $a_2+c_1$ & $e_1$ & $f_1$ & $c_4+f_1$\\
    & $b_2+c_2$ & & & $e_2+f_2$\\
    \hline
    \multirow{3}{*}{$\theta =4$} & $a_1$ & $a_1+d_1$ & $b_1+d_3$ & $e_2$ \\
    & $b_1$ & $a_2+e_1$ & $b_2+f_1$ & $f_2$\\
    &  & $d_2+e_2$ & $d_4+f_2$ &\\
    \hline
     \multirow{3}{*}{$\theta =5$} & $a_2$ & $a_1+d_1$ & $d_2$ & $c_1+e_3$ \\
    & $c_1$ & $a_2+e_1$ & $f_2$ & $c_2+f_1$\\
    &  & $d_2+e_2$& & $e_4+f_2$\\
    \hline
    \multirow{3}{*}{$\theta =6$} & $b_2$ & $d_2$ & $b_1+d_1$ & $c_1+e_1$ \\
    & $c_2$ & $e_2$ & $b_2+f_1$ & $c_2+f_3$\\
    &  & & $d_2+f_2$& $e_2+f_4$\\
    \hline
    \end{tabular}
    \caption{Retrieval scheme for $\mathbf{K}_4$ in Example~\ref{ex_complete}.}
    \label{tab:K4 answers}
\end{table} 

Table~\ref{tab:K4 answers} illustrates the scheme for local PIR for each required message index. To retrieve $4$ desired message symbols, we download $10$ symbols in total, yielding the rate $\frac{4}{10}=0.4$. As shown in \cite{gePIR}, the PIR capacity of $\mathbf{K}_4$ is bounded as $0.35\leq C_{PIR}(\mathbf{K}_4)\leq 0.3529$. Thus, the rate of the local PIR scheme for $\mathbf{K}_4$ is higher than the upper bound on the PIR capacity. 
\end{example}

These examples demonstrate the potential rate improvements in local PIR, compared to standard PIR, due to the relaxed notion of local privacy. Next, we recall some useful definitions for graphs that will be used in our analysis.

\subsection{Definitions}

\begin{definition}[Disjoint union of graphs] 
    Graph $G=(V,E)$ is a disjoint union of $m$ graphs $G_i = (V_i,E_i)$, $i\in[m]$, denoted by $G = \cup_{i \in [m]} G_i$, if
    \begin{align}
        &V_i \cap V_j = \emptyset, \quad i,j\in [m],~ i\neq j,\\
        &V=\bigcup_{i=1}^m  V_i,  \quad E = \bigcup_{i=1}^m E_i.
    \end{align}
\end{definition}

This definition is required to connect the local PIR rate of $G$ to that of each individual subgraphs $G_i, ~ i\in [m]$ (see Theorem~\ref{thm:disjoint_graphs_capacity}).

\begin{definition}[Bipartite graphs]\label{def_bipartite}
    Graph $G=(V,E)$ is bipartite, if the vertex set $V$ is a disjoint union of two sets $V_1$ and $V_2$, and every edge $\{i,j\}\in E$, has $i\in V_1$ and $j\in V_2$. These two sets $V_1$ and $V_2$ are called parts.
\end{definition}

Examples of bipartite graphs are cyclic graphs with an even number of nodes, path graphs, star graphs and complete bipartite graphs. Given a graph $G=(V,E)$, its \emph{vertex cover} is a subset of vertices $V'\subseteq V$ such that every edge in $E$ has at least one end point in $V'$. Note that, in Definition \ref{def_bipartite} each of the parts $V_1$ and $V_2$ is a vertex cover of $G$. Individually, $V_1$ and $V_2$ are also \emph{independent sets} of $G$, i.e., the nodes in $V_1$ (or $V_2$) share no common edge among each other. This special property of bipartite graphs helps us design an efficient local PIR scheme for such graphs (see Theorem~\ref{thm:ach_bipartite}). 

\begin{definition}[Edge-transitive graphs]\label{def:edge_trans}
    Graph $G=(V,E)$ is edge-transitive, if for every pair of edges $e_1,e_2\in E$ there exists an automorphism\footnote{An \emph{automorphism} of graph $G$ is a permutation $\sigma$ of the vertices in $V$, which preserves the graph structure in terms of vertex connectivity. That is, for any $u,v\in V$, $\sigma(u)$ and $\sigma(v)$ share an edge if and only if $u$ and $v$ share an edge. } that maps $e_1$ to $e_2$. 
\end{definition}

Examples of edge-transitive graphs are cyclic graphs, star graphs, complete graphs and complete bipartite graphs. The symmetry of such a graph $G$ allows us to exactly define a local PIR scheme for $G$ by a scheme on a single edge (see Theorem~\ref{thm:edge_trans}).\\

\section{Main Results for Simple Graphs}\label{sec:simple graph results}
In this section, we summarize the results for simple graphs, i.e., graphs where each message is replicated exactly twice, and no more than one edge exists between any two distinct nodes. Subsequently, we use the term graph instead of simple graph. 

The first result connects the local PIR capacity of graphs to the local PIR capacity of their disjoint union. Without loss of generality, we assume the individual graphs to be connected, i.e., there exists a path between every pair of vertices.

\begin{theorem}\label{thm:disjoint_graphs_capacity}
    Let $G = \bigcup_{i=1}^m G_i$, where $G_i = (V_i, E_i)$. Let the capacity-achieving local PIR scheme on $G_i$ entail the message length $L_i$, and the expected download cost $\mathbb{E}[D_i]$  for the respective $|E_i|$ messages. Then, the local PIR capacity of $G$ is given by,
    \begin{align}\label{eq:disjoint_union_graphs}
        C(G)  = \frac{\sum_{i=1}^m  |E_i| \cdot L_i}{\sum_{i=1}^m |E_i| \cdot \mathbb{E}[D_i]}.
    \end{align}
\end{theorem}

\begin{Proof}
We present the proof for $m=2$, and the case of $m>2$ follows by induction. Let $E_1=[K_1]$ and $E_2=[K_1+1:K_1+K_2]$. Then, the scheme $\Pi$ can be described as follows. Given $k\in E_1\cup E_2$, the user follows the optimal local PIR scheme on $G_1$ if $k\in E_1$ and the optimal local PIR scheme on $G_2$ if $k\in E_2$, which yields the rate in \eqref{eq:disjoint_union_graphs}. Suppose, for the sake of contradiction, that there exists a scheme $\Pi'$ that achieves $R^{\Pi'}(G)> \frac{|E_1|L_1+|E_2|L_2}{|E_1|\mathbb{E}[D_1] + |E_2|\mathbb{E}[D_2]}$. Then, if $V_2=\emptyset$, we have $G=G_1$, and that $C(G)\geq R^{\Pi'}(G)>\frac{L_1}{\mathbb{E}[D_1]}$. This leads to a contradiction, since $C(G)=C(G_1)$ is at most $\frac{L_1}{\mathbb{E}[D_1]}$, hence, no such $\Pi'$ exists.   
\end{Proof}

\begin{corollary}
    If the $m$ graphs $G_1,\ldots,G_m$ are identical, then $C(G) = C(G_1)$, i.e., there is no reduction in capacity. This provides multiplicative gain compared to the PIR capacity in the same setting, since $C(G)\geq C_{PIR}(G) = \frac{1}{m}C_{PIR}(G_1)$.
\end{corollary}

The subsequent results are for connected graphs. For each $k\in [K]$, let $i$ and $j$ be the two servers where $W_k$ is replicated, i.e., $\{k\}= \mathcal{I}_i \cap \mathcal{I}_j$. Consider the subgraph $G_k=(V_k,E_k) \subseteq (V,E) = G$ with
\begin{align}
    V_k &=\{m\in [N]:  \mathcal{I}_m\cap(\mathcal{I}_i\cup \mathcal{I}_j)\neq \emptyset\}\subseteq V, \label{def_v_k}\\
    E_k &= \{\ell \in [K]:\ell \in \mathcal{I}_i\cup \mathcal{I}_j\}\subseteq E, \label{def_e_k}
\end{align}
i.e., $V_k$ is the set of servers that share at least one message with server $i$ or server $j$, including servers $i$ and $j$ and $E_k$ is the set of messages that are stored at server $i$ or server $j$. If $G$ is edge-transitive, $G_k$ is identical for all $k\in [K]$. Examples are cyclic, star, complete, complete bipartite and Petersen graphs.

\begin{theorem}\label{thm:edge_trans}
    For an edge-transitive graph $G$, fix the edge $\{i,j\}$ and the corresponding sub-graph $G_k$. Then, the local PIR capacity is lower bounded as 
    \begin{align}\label{eq:thm 2_gen}
        C(G)\geq \max_{t_i\in [\deg(i)], t_j\in [\deg(j)]} \ \frac{1}{\lambda(t_i,t_j)\left(\frac{\deg(i)}{t_i}+t_i-1\right)+(1-\lambda(t_i,t_j))\left(\frac{\deg(j)}{t_j}+t_j-1\right)}, 
    \end{align}
    where
    \begin{align}
        \lambda(t_i,t_j) = \frac{\binom{\deg(i)-1}{t_i-1}}{\binom{\deg(i)-1}{t_i-1}+\binom{\deg(j)-1}{t_j-1}}.
    \end{align}
\end{theorem}

\begin{remark}\label{rem:equal deg i and j}
    If $\deg(i)=\deg(j)=d$ for all $G_k$, then \eqref{eq:thm 2_gen} can be simplified to 
    \begin{align}
        C(G)&\geq 
        \max_{t\in [d]}\frac{1}{\left(\frac{d}{t}+t-1\right)}\\
        &= \max\left\{\frac{1}{\frac{d}{\lfloor\sqrt{d}\rfloor}+\lfloor\sqrt{d}\rfloor-1},\frac{1}{\frac{d}{\lceil\sqrt{d}\rceil}+\lceil\sqrt{d}\rceil-1}\right\}\\
        &\geq \frac{1}{2\sqrt{d}}.\label{eq:remark_final ineq}
    \end{align}
\end{remark}

The proofs of Theorem~\ref{thm:edge_trans} and Remark~\ref{rem:equal deg i and j} appear in Section~\ref{proof:edge_trans} and Section~\ref{proof:edge_trans_equal_deg}, respectively.

\begin{corollary}\label{cor_2}
    From Theorem~\ref{thm:edge_trans}, we obtain the following rates for common graph families.
    \begin{enumerate}
    \item For the cyclic graph $G=\mathbf{C}_N$, with $N\geq 4$ the subgraph $G_k=\mathbf{P}_4$ with $\deg(i) = \deg(j) = 2$, this yields
    \begin{align}\label{eq:lbnd_cycle}
        C(\mathbf{C}_N)\geq \frac{1}{2},
    \end{align}  
    which is equal to $C_{PIR}(\mathbf{C}_N)=\frac{2}{N+1}$ if $N=3$, and strictly greater for all $N \geq 4$. Interestingly, the rate $\frac{1}{2}$ for all $N$ also coincides with the standard PIR capacity of the subgraph $\mathbf{P}_4$ (where $C_{PIR}(\mathbf{P}_N)=\frac{2}{N}$).\label{cycle_rate}
    
    \item For the complete graph $G=\mathbf{K}_N$, $\deg(i) = \deg(j) = N-1$, this yields 
    \begin{align}
     C(\mathbf{K}_N)\geq
     \begin{cases}
     \frac{1}{2\sqrt{N-1}-1}, & \text{ $N-1$ perfect square,}\\
         \frac{1}{2\sqrt{N-1}}, & \text{ otherwise,}
     \end{cases}
    \end{align} 
    while $\left(\frac{4}{3}-o(1)\right)\frac{1}{N}\leq C_{PIR}(\mathbf{K}_N)\leq \frac{1}{N(e-2)}$, as recently established in \cite{gePIR}.\label{complete_rate}
    
    \item For the complete, balanced bipartite graph $G=\mathbf{K}_{\frac{N}{2},\frac{N}{2}}$ for even $N$, $\deg(i) = \deg(j) = \frac{N}{2}$, this yields
    \begin{align}\label{eq:bipartite_balanced}
        C(\mathbf{K}_{\frac{N}{2},\frac{N}{2}})\geq 
         \begin{cases}
         \frac{1}{\sqrt{2N}-1}, & \text{ $\frac{N}{2}$ perfect square,}\\
             \frac{1}{\sqrt{2N}}, & \text{ otherwise,}
         \end{cases}
    \end{align}
    while $\frac{4}{3N}\leq C_{PIR}(\mathbf{K}_{\frac{N}{2},\frac{N}{2}})\leq \frac{1}{N(e^{0.5}-1)}$, by the tightest lower and upper bounds, established in \cite{krishnan_graph} and \cite{gePIR}, respectively.  
    \end{enumerate}
\end{corollary}

Next, we establish the local PIR rate for a general graph $G$, which is derived from the scheme of Theorem~\ref{thm:edge_trans} for $\mathbf{K}_N$. This relies on the fact that $G$ is a sub-graph of $\mathbf{K}_N$ for any connected graph $G$ with $N$ vertices. 

\begin{theorem}\label{thm:gen_ach_complete_based}  
    Consider the row vector $\bm{\delta} =[ \delta_1, \ldots, \delta_N ]$, where  $\delta_n = (N-1)-\deg(n)$, $n\in [N]$. Then, for any graph $G=(V,E)$ with $|V|=N$, $|E|=K$, and $L=2\binom{N-2}{t-1}$, the local PIR capacity is lower-bounded as
    \begin{align}\label{eq:rate expression_gen_comp}
        C(G)\geq \max_{t \in [N-1]} \ \frac{1}{\frac{1}{R^{\Pi_t}(\mathbf{K}_N)}-\frac{\sum_{k=1}^K\Delta_k}{KL}},
    \end{align}
    where, for every $\{k\}=\mathcal{I}_i\cap \mathcal{I}_j$, with $i,j\in [N]$,
    \begin{align}
        \Delta_k = \binom{\delta_i}{t}+\binom{\delta_j}{t}+(\delta_i+\delta_j)\binom{N-3}{t-2},
    \end{align}
    and $\Pi_t$ is the scheme for edge-transitive graphs in Theorem~\ref{thm:edge_trans} for a given $t$.
\end{theorem}

\begin{corollary} \label{m_cor_3}
    From Theorem~\ref{thm:gen_ach_complete_based}, we obtain the following lower bounds.
    \begin{enumerate} 
    \item For $\mathbf{C}_N$ and $\mathbf{K}_N$, we recover the same rates as in Corollary~\ref{cor_2}.\ref{cycle_rate} and \ref{cor_2}.\ref{complete_rate}, respectively.
    
    \item For the path graph $G = \mathbf{P}_N$, $\bm{\delta}=[N-2, \underbrace{N-3,\ldots,N-3}_{N-2 \text{ times}},N-2]$, and we have the capacity lower bound
    \begin{align}
        C(\mathbf{P}_N) \geq &  \max_{t \in [N-1]}\frac{1}{\frac{N-1}{t}+t-1 - \frac{(N-t-1)^2/t +(N-2)(t-1)}{N-1}}\\
        =& \frac{N-1}{2N-3}.
    \end{align}
      
    \item For the wheel graph $G = \mathbf{W}_N$, where server $1$ is the central node, $\bm{\delta} = [0, N-4, \ldots, N-4]$. Consequently, we obtain $C(\mathbf{W}_{N}) = \Omega\left(\frac{1}{\sqrt{N}}\right)$ for asymptotically large $N$. This bound is larger than the upper bound for wheel graphs in the PIR setting, where the upper bound is $\frac{1}{N}$ \cite{SGT23}.
    \end{enumerate}
\end{corollary}  

\begin{remark}
    For the complete graph $\mathbf{K}_N$ and the complete, balanced bipartite graphs $\mathbf{K}_{\frac{N}{2},\frac{N}{2}}$, the local PIR capacity is $\Omega\left(\frac{1}{\sqrt{N}}\right)$, whereas the standard PIR capacity for these graphs is $\Theta\left(\frac{1}{N}\right)$ as $N\to \infty$. That is, the PIR capacity vanishes to $0$ approximately $\sqrt{N}$ times faster.
\end{remark}

\begin{remark}
For the star graph $G=\mathbf{S}_N$, the subgraph $G_k=G$; the lower bounds on $C(\mathbf{S}_N)$ yielded by Theorem~\ref{thm:edge_trans} and Theorem~\ref{thm:gen_ach_complete_based} are $\Omega\left(\frac{1}{\sqrt{N}}\right) $ (see Table~\ref{tab:all_rates}), which coincide with the PIR capacity for asymptotically large $N$ \cite{SGT23,  YJ23}. However, leveraging the fact that $\mathbf{S}_N=\mathbf{K}_{N-1,1}$, the next theorem for the class of bipartite graphs improves this, and identifies the local PIR capacity for $\mathbf{S}_N$.
\end{remark}

\begin{theorem}\label{thm:ach_bipartite}
    If $G=(V,E)$ is bipartite, with parts $V_1,V_2$, the local PIR capacity is lower bounded as
    \begin{align}
            C(G)\geq K\cdot \left(\min_{m\in \{1,2\}}\sum_{n\in V_m } \deg(n)^2\right)^{-1}.\label{eq:lbnd_bipartite}
    \end{align}
\end{theorem}

The proof allows for minimal subpacketization, i.e., $L=1$ and follows from the scheme in Section~\ref{proof:bipartite}.

\begin{corollary}\label{cor_3}
    From Theorem~\ref{thm:ach_bipartite}, the following rates for common graph families are achievable.
    \begin{enumerate}
        \item For the star graph $G=\mathbf{S}_N$, this yields
        \begin{align}
            C(\mathbf{S}_{N}) = 1,
        \end{align}
        by choosing $V_1=\{1,\ldots,N-1\}$ and $V_2=\{N\}$, since $\deg(n)=1$ for each $n\in V_1$, and $K=N-1$. This is the capacity since trivially, $C(G)\leq 1$ for any $G$.
        
        \item For the path graph $\mathbf{P}_N$, we obtain
        \begin{align}\label{eq:path_lbnd}
            C(\mathbf{P}_N)\geq \begin{cases}
                \frac{N-1}{2N-3}, & N \text{ even},\\
                \frac{N-1}{2N-4}, & N \text{ odd},
            \end{cases}
        \end{align}
         by choosing $V_1$ as odd and $V_2$ as even vertex indices in $[N]$. The lower bound here for even $N$ matches the lower bound in Corollary \ref{m_cor_3}.2, while it is improved for odd $N$. \label{cor:path}
         
        \item For the cyclic graph $\mathbf{C}_N$ with even $N$, the rate is $\frac{1}{2}$, equal to \eqref{eq:lbnd_cycle}.
        \item For complete bipartite graphs $\mathbf{K}_{N_1,N_2}$, we have $C(\mathbf{K}_{N_1,N_2})\geq \max\left\{\frac{1}{N_1},\frac{1}{N_2}\right\}$, which for the balanced case of $\mathbf{K}_{\frac{N}{2},\frac{N}{2}}$ with even $N$ reduces to 
        \begin{align}
            C(\mathbf{K}_{\frac{N}{2},\frac{N}{2}}) \geq \frac{2}{N}.
        \end{align}
        This is strictly better than the bound \eqref{eq:bipartite_balanced} for $N\leq 6$, and is lower otherwise. 
    \end{enumerate} 
\end{corollary}

The following theorem establishes another capacity lower bound for any connected graph $G$. Similar to that of Theorem~\ref{thm:ach_bipartite}, the achievable scheme requires minimal subpacketization, i.e., $L=1$, and is presented in Section~\ref{proof_L=1}. 

\begin{theorem}\label{thm:gen_graph_ach}
    For any graph $G=(V,E)$ the capacity of local PIR is lower-bounded as
    \begin{align}
        C(G)\geq 2K\cdot \left(\sum_{n\in V } \deg(n)^2\right)^{-1}.\label{eq:lbnd_general}
    \end{align}
\end{theorem}

\begin{remark}
    It can be verified that, if $G$ is bipartite, the rate provided by Theorem \ref{thm:ach_bipartite} is greater than or equal to the rate from Theorem \ref{thm:gen_graph_ach}, for the same subpacketization $L=1$.
\end{remark}

\begin{table}[t]
    \centering
    \renewcommand{\arraystretch}{1.35}
    \setlength{\tabcolsep}{2pt}
    \begin{tabular}{|c|c|c|c|c||c|}
    \hline
    \textbf{Graph type} & \textbf{Thm.~\ref{thm:edge_trans}} & \textbf{Thm.~\ref{thm:gen_ach_complete_based}} & \textbf{Thm.~\ref{thm:ach_bipartite}} & \textbf{Thm.~\ref{thm:gen_graph_ach}} & $C_{PIR}$\\
    \hline
    Path $\mathbf{P}_N$ & - & $\frac{N-1}{2N-3}$ & $\begin{cases}\frac{N-1}{2N-3}, & N \text{ even}\\
        \frac{N-1}{2N-4}, & N \text{ odd}
        \end{cases}$ & $\frac{N-1}{2N-3}$ & $\frac{2}{N}$ \\
    \hline
    Cyclic $\mathbf{C}_N$ & $\frac{1}{2}$& $\frac{1}{2}$ &  $\frac{1}{2},  ~N$ even & $\frac{1}{2}$ & $\frac{2}{N+1}$\\
    \hline
    Complete $\mathbf{K}_N$ & 
    $\frac{1}{2\sqrt{N-1}}
    $ & $\frac{1}{2\sqrt{N-1}}
    $ & - & $\frac{1}{N-1}$ & $\Theta\left(\frac{1}{N}\right)$ \\
    \hline
    Star $\mathbf{S}_N$ & $\frac{1}{2\sqrt{N-1}}
    $ & $\frac{1}{\sqrt{N-1}}
    $ & $1$ & $\frac{2}{N}$ & $\Theta\left(\frac{1}{\sqrt{N}}\right)$\\
    \hline
    Complete, balanced bipartite $\mathbf{K}_{\frac{N}{2},\frac{N}{2}}$ & $ \frac{1}{\sqrt{2N}}$ &  $ \frac{1}{\sqrt{2N}}$ & $\frac{2}{N}$ & $\frac{2}{N}$ & $\Theta\left(\frac{1}{N}\right)$\\
    \hline
    Wheel $\mathbf{W}_N$ & - & $\Omega\left(\frac{1}{\sqrt{N}}\right)$ & - &$\frac{4}{N-8}$ & $\Theta\left(\frac{1}{N}\right)$\\
    \hline
    \end{tabular}
    \caption{Achievable local PIR rates and exact/approximate PIR capacities for common graph families. A hyphen (-) indicates that the result does not apply to the graph.}
    \label{tab:all_rates}
\end{table}

Table~\ref{tab:all_rates} provides a summary of local PIR rates for common graph families, alongside the corresponding PIR capacities. In all cases, the greatest achievable rates for local PIR are superior to the respective PIR capacities. 

\begin{theorem}\label{thm:cyclic_converse}
    For the cyclic graph $\mathbf{C}_N$, the local PIR capacity is given by
    \begin{align}
      C(\mathbf{C}_N)=\frac{1}{2}.  
    \end{align}
\end{theorem}

\begin{theorem}\label{thm:path_converse}
    For the path graph $\mathbf{P}_N$, the local PIR capacity satisfies
    \begin{align}
        C(\mathbf{P}_N)\leq \frac{N-1}{2N-4}.
    \end{align}
    This matches the rate \eqref{eq:path_lbnd} for odd $N$, which gives $C(\mathbf{P}_N)=\frac{N-1}{2N-4}$ for odd $N$.
\end{theorem}

\begin{remark}
    Interestingly, the local PIR capacities of path, cyclic and star graphs are strictly bounded above $0$ as $N\to \infty$. This highlights an important gain in the communication efficiency compared to the standard PIR model for the aforementioned graphs, where the capacity is known to vanish to $0$ as $N\to \infty$. 
\end{remark}

\begin{remark}\label{rmk:half_upper_bnd}
    For any graph $G$ with $\deg(n)\geq 2$ for all vertices $n\in[N]$, the capacity upper-bound
    $C(G)\leq \frac{1}{2}$
    holds in general.
\end{remark} 

The proof follows from a simple observation and is shown in Section~\ref{pf:half_upper_bnd}. As a corollary, local PIR capacity on $\mathbf{K}_N, \mathbf{K}_{N_1,N_2}~(N_1+N_2=N)$ and $\mathbf{W}_N$ satisfy this, besides $\mathbf{C}_N$.

\section{Proofs of Capacity Lower Bounds}\label{sec:ach proofs}
In this section, we present the achievable schemes that yield the rates in Theorems~\ref{thm:edge_trans}-\ref{thm:gen_graph_ach}. Each scheme description is followed by a running example on the complete bipartite graph $\mathbf{K}_{2,3}$ (see Fig.~\ref{fig:K_2,3}), where $N=5$ and $K=6$. Without loss of generality, we compute the rates only for $\theta=1$, which applies to all indices by the edge-transitivity of $\mathbf{K}_{2,3}$.

\subsection{Proof of Theorem~\ref{thm:edge_trans}}\label{proof:edge_trans}
We present the scheme for a fixed $G_k$ where $\{k\}=\mathcal{I}_i\cap \mathcal{I}_j$ when $\theta = k$. By edge-transitivity of $G$, the same scheme applies to all $G_k$. The user communicates only with the servers in $V_k$ and queries only for the messages corresponding to $E_k$, i.e., $\mathcal{W}_i\cup \mathcal{W}_j$. We fix $t_i, t_j\in \mathbb{N}$, with $t_i \in [\deg(i)]$ and $t_j \in [\deg(j)]$. The scheme requires each message to consist of 
\begin{align}\label{eq:subpacketization}
    L=\binom{\deg(i)-1}{t_i-1}+\binom{\deg(j)-1}{t_j -1} 
\end{align}
symbols. The user applies a private, independent permutation to the message symbols of each message in $\mathcal{W}_i\cup \mathcal{W}_j$, with $W_\ell(m)$ denoting the $m$th permuted symbol of $W_\ell$. To servers $i$ and $j$, the user sends queries for all possible $t_i$-sums\footnote{The term $l$-sum for $l\geq 1$ refers to the sum of symbols of $l$ distinct messages.} of the messages in $\mathcal{W}_i$, and all possible $t_j$-sums of the messages in $\mathcal{W}_j$, respectively. In particular, let $\mathcal{A}_i$ denote the ordered set of all $t_i$-subsets of $\mathcal{I}_i$,  and $\mathcal{A}_j$ denote the ordered set of all $t_j$-subsets of $\mathcal{I}_j$ arranged in a lexicographic order. Let $\mathcal{A}_i=\big(A^i_1,\ldots, A^i_{\binom{\deg(i)}{t_i}}\big)$ and $\mathcal{A}_j=\big(A^j_1,\ldots, A^j_{\binom{\deg(j)}{t_j}}\big)$. Then, for each $\ell \in \mathcal{I}_i$ and each $A_p^i\in \mathcal{A}_i$, such that $\ell\in A_p^i$, we define $\gamma_\ell(A_p^i)$ as
\begin{align}\label{eq:gamma_define}
    \gamma_\ell(A_p^i) = \sum_{x=1}^{p}{\mathds{1}\{\ell\in A^i_x\}},
\end{align}
i.e., the number of times that $\ell$ is contained in a subset, up to and including $A_p^i$. The same definition of $\gamma_{\ell}(A_p^{j})$ extends to $\mathcal{A}_j$. For each $A\in \mathcal{A}_i$, the user downloads from server $i$,
\begin{align}\label{eq:download server i}
    \sum_{\ell\in A} W_\ell(\gamma_\ell(A)).
\end{align}
From server $j$, for each  $A\in\mathcal{A}_j$, the user downloads 
\begin{align}\label{eq:download server j}
    \begin{cases}
        W_k\left(\binom{\deg(i)-1}{t_i-1}+\gamma_k(A)\right)+\sum_{\ell\in A\setminus\{k\}}W_\ell(\gamma_\ell(A)),& k\in A,\\
      \sum_{\ell\in A} W_\ell(\gamma_\ell(A)),& k\notin A.
    \end{cases} 
\end{align}
For each $\ell\in \mathcal{I}_i$, and each $\ell\in \mathcal{I}_j$, $\gamma_\ell(A)$ varies from 1 to $ \binom{\deg(i)-1}{t_i-1}$, and from 1 to $ \binom{\deg(j)-1}{t_j-1}$, respectively. From the servers, $V_k \setminus \{i,j\}$, the user downloads the interference message symbols, that appear in the summations \eqref{eq:download server i} and the first case of \eqref{eq:download server j}. Specifically, for each $\ell\in \mathcal{I}_i\setminus\{k\}$, define
\begin{align}\label{B_define1}
    \mathcal{B}_{\ell,i} =\{B\in \mathcal{A}_i: \{\ell,k\}\subseteq B\},
\end{align}
and for each $\ell\in \mathcal{I}_j\setminus \{k\}$, define
\begin{align}\label{B_define2}
    \mathcal{B}_{\ell,j} =\{B\in \mathcal{A}_j: \{\ell,k\}\subseteq B\}.
\end{align}
The sets $\mathcal{B}_{\ell,i}$ and $\mathcal{B}_{\ell,j}$ consist of the indices of $W_\ell$ appearing in \eqref{eq:download server i} and \eqref{eq:download server j}, respectively. Then, corresponding to $\ell\in \mathcal{I}_i\setminus \{k\}$ (or $\mathcal{I}_j\setminus\{k\}$), the user downloads the symbols,
\begin{align}\label{eq:download from remaining servers}
    \{W_\ell(\gamma_\ell(B)): B\in \mathcal{B}_{\ell,i} \text{ (or $  \mathcal{B}_{\ell,j}$)}\}
\end{align}
from the server storing $W_\ell$. Since the graph is simple, a single server in $V_k\setminus \{i,j\}$ stores $W_\ell$. The number of downloaded symbols to retrieve the $L$ symbols of $W_k$ is
\begin{align}
    D_k &=|\mathcal{A}_i| + |\mathcal{A}_j| + \sum_{\ell\in \mathcal{I}_i\setminus \{k\}}|\mathcal{B}_{\ell,i}|+ \sum_{\ell\in \mathcal{I}_j\setminus \{k\}}|\mathcal{B}_{\ell,j}| \\
    &=\binom{\deg(i)}{t_i}+\binom{\deg(j)}{t_j}+(\deg(i)-1)\binom{\deg(i)-2}{t_i-2}+(\deg(j)-1)\binom{\deg(j)-2}{t_j-2}\\
    &=\binom{\deg(i)-1}{t_i-1}\left(\frac{\deg(i)}{t_i}+t_i-1\right) + \binom{\deg(j)-1}{t_j-1}\left(\frac{\deg(j)}{t_j}+t_j-1\right).\label{eq:download_cost} 
\end{align}
The ratio of \eqref{eq:subpacketization} and \eqref{eq:download_cost} gives the desired rate. The rate is maximized over all choices of $t_i\in [\deg(i)]$ and $t_j\in [\deg(j)]$ to obtain the largest lower bound. 

For $\theta = k$, privacy holds at the servers $i$ and $j$, as required. This is because, once the scheme parameters $t_i$ and $t_j$ are fixed, the same query structure is maintained for all messages stored at server $i$ (and server $j$) for all $\theta$ in $\mathcal{I}_i$ (and $\mathcal{I}_j)$ due to edge-transitivity of the graph. Moreover, the servers remain oblivious of the actual queried message symbols, since the symbol indices are permuted privately and independently by the user.

\begin{table}[t]
    \centering
    \begin{tabular}{|c|c|c|c|c|}
    \hline
     server 1 & server 2& server 3& server 4& server 5\\
    \hline
    $W_1$ & $W_4$ & $W_1$ & $W_2$ & $W_3$\\
    $W_2$ & $W_5$ & $W_4$ & $W_5$ & $W_6$\\
    $W_3$ & $W_6$ & & & \\
    \hline
    \end{tabular}
    \caption{Storage for $\mathbf{K}_{2,3}$; see the graph in Fig.~\ref{fig:K_2,3}.}
    \label{tab:ex_bipartite_storage}
\end{table}

Examples~\ref{ex_cycle} and \ref{ex_complete} in Section~\ref{motive_ex} follow this scheme. The next example is on $\mathbf{K}_{2,3}$ with storage given in Table~\ref{tab:ex_bipartite_storage}; see the graph in Fig.~\ref{fig:K_2,3}.

\begin{example}\label{ex:edge_trans_bip}
The optimal choices of the optimization variables in \eqref{eq:thm 2_gen} are $t_i = 2$ and $t_j = 1$. Let $\theta=1$, then $i=1$, and $j=3$ with $V_1 = V=[5]$, $E_1 = \{1,2,3,4\}$ and $L=3$. The sets $\mathcal{A}_n$ and $\mathcal{B}_{\ell,n}$ for $n=1,3$ and $\ell \in \{2,3,4\}$ are
\begin{align}
    \mathcal{A}_1 &= \left(\{1,2\},\{1,3\},\{2,3\}\right), \quad \mathcal{A}_3 = \left(\{1\},\{4\}\right),\\
    &\mathcal{B}_{2,1} = \{\{1,2\}\}, \quad \mathcal{B}_{3,1} = \{\{1,3\}\},  \quad \mathcal{B}_{4,2} =\emptyset.
\end{align}
From here, we deduce the corresponding answers as given in Table~\ref{tab:answers_bip_1}, and the rate is $\frac{3}{7}$.
\end{example}

\begin{table}[t]
    \centering
    \begin{tabular}{|c|c|c|c|c|}
    \hline
        server 1 & server 2 & server 3 & server 4 & server 5\\
        \hline
    $a_1+b_1$ & $\emptyset$ & $a_3,d_1$ & $b_1$ & $c_1$\\
    $a_2+c_1$ & & & &\\
    $b_2+c_2$ & & & &\\
    \hline
    \end{tabular}
    \caption{Retrieval scheme for $\mathbf{K}_{2,3}$ for Example~\ref{ex:edge_trans_bip}.}
    \label{tab:answers_bip_1}
\end{table}

\subsection{Proof of Remark~\ref{rem:equal deg i and j}}\label{proof:edge_trans_equal_deg}
In the rate expression, we minimize the term
\begin{align}\label{eq:convex_comb}
    & \lambda(t_i,t_j)\left(\frac{d}{t_i}+t_i-1\right)+(1-\lambda(t_i,t_j))\left(\frac{d}{t_j}+t_j -1\right),
\end{align}
where 
\begin{align}
\lambda(t_i,t_j) & = \frac{\binom{d-1}{t_i-1}}{\binom{d-1}{t_i-1}+\binom{d-1}{t_j-1}}\in (0,1).
\end{align}
Assume without loss of generality, that $\frac{d}{t_i}+t_i-1\leq \frac{d}{t_j}+t_j -1$. Now, for any $\lambda(t_i,t_j)\in (0,1)$, 
\begin{align}
    \lambda(t_i,t_j)\left(\frac{d}{t_i}+t_i-1\right)+(1-\lambda(t_i,t_j))\left(\frac{d}{t_j}+t_j -1\right)&\geq \min\left(\frac{d}{t_i}+t_i-1, \frac{d}{t_j}+t_j-1\right)\\
    &= \frac{d}{t_i}+t_i-1,\label{eq:denominator}
\end{align}
The minimum value of \eqref{eq:denominator} is achieved upon setting $t_i = t_j = t $ in \eqref{eq:convex_comb}. This simplifies the problem to
\begin{align}
     \min_{t\in [N-1]} \frac{d}{t}+t-1,
\end{align}
which is solved by $t=\lfloor\sqrt{d}\rfloor$ or $t=\lceil \sqrt{d}\rceil$, or both if both values yield the same ${\frac{d}{t}+t-1}$. Finally, the last inequality \eqref{eq:remark_final ineq} follows since $\frac{d}{\lfloor\sqrt{d}\rfloor}-1$, $\lfloor\sqrt{d}\rfloor$, $\frac{d}{\lceil\sqrt{d}\rceil}$ and $\lceil{\sqrt{d}}\rceil-1$  are all upper-bounded by $\sqrt{d}$.

\subsection{Proof of Theorem~\ref{thm:gen_ach_complete_based}}\label{proof_complete_based}
First, we fix $t\in [N-1]$ and $L = 2\binom{N-2}{t-1}$. Let the user privately permute the symbols of all messages in $\mathcal{W}=\{W_1,\ldots,W_K\}$, by picking a permutation uniformly at random, from the set of all permutations of $L$ permutations, and independently for each $W_k\in \mathcal{W}$. Consider the \emph{helper} complete graph $\mathbf{K}_N$ with the message set $\mathcal{W}^{\mathbf{K}_N} \supset  \mathcal{W}$ where $W_{K+1}, \ldots, W_{\binom{N}{2}}$ are the virtual messages, i.e., messages that are missing in the storage corresponding to $G$. Let $\mathcal{I}_n^{\mathbf{K}_N}$ denote the set of message indices stored in server $n\in[N]$ of the helper $\mathbf{K}_N$. With this setup, we describe the local PIR scheme for $\theta = k$ where $\mathcal{I}_i\cap \mathcal{I}_j = \{k\}$. By recalling the definitions of $V_k$ in \eqref{def_v_k} and $E_k$ in \eqref{def_e_k}, the user interacts only with the servers corresponding to $V_k$, and the answers consist of only the messages corresponding to $E_k$, i.e., $\mathcal{W}_i\cup \mathcal{W}_j$.

Following the scheme for edge-transitive graphs in Section~\ref{proof:edge_trans}, we construct the ordered sets $\mathcal{A}_i^{\mathbf{K}_N}$ and $\mathcal{A}_j^{\mathbf{K}_N}$ using the index sets $\mathcal{I}_i^{\mathbf{K}_N}$ and $\mathcal{I}_j^{\mathbf{K}_N}$, respectively. From here, the user constructs the ordered multisets $\mathcal{A}_i$ and $\mathcal{A}_j$ as follows. For each $A\in \mathcal{A}_i^{\mathbf{K}_N}$, we remove the elements in $\mathcal{I}_i^{\mathbf{K}_N}\setminus \mathcal{I}_i$, and for each $A\in \mathcal{A}_j^{\mathbf{K}_N}$, we remove the elements in $\mathcal{I}_j^{\mathbf{K}_N}\setminus \mathcal{I}_j$. Then, for a fixed $t$, $\mathcal{A}_i$ and $\mathcal{A}_j$ consist of the resultant non-empty subsets of $\mathcal{I}_i$ and $\mathcal{I}_j$, respectively, of size at most $t$. From server $i$, the user queries for sums involving the first (permuted) $L/2$ symbols of $W_k$. Similarly, from server $j$, the user queries for sums involving the last (permuted) $L/2$ symbols of $W_k$. More specifically, for each $A\in \mathcal{A}_i$ the user queries for the sum given by \eqref{eq:download server i}. Similarly, for each $A\in \mathcal{A}_j$, the user queries for the sum given by \eqref{eq:download server j} where $\binom{\deg(i)-1}{t_i-1} = \binom{N-2}{t-1}$. Note that, for each $\ell$, $\gamma_\ell(A)$ varies from $1$ to $\binom{N-2}{t-1}$ for $A\in \mathcal{A}_i$ or $A \in \mathcal{A}_j$. To account for the interfering message symbols, we construct the multisets $\mathcal{B}_{\ell,i}$ and $\mathcal{B}_{\ell,{j}}$, following \eqref{B_define1}-\eqref{B_define2}. The user downloads the interfering message symbols given by \eqref{eq:download from remaining servers} from the remaining servers $V_k\setminus \{i,j\}$.

Let $D^{\mathbf{K}_N}$ be the local PIR download cost for the $\mathbf{K}_N$ storage, which is independent of $\theta=k$, by edge-transitivity of $\mathbf{K}_N$. The download cost is given by
\begin{align}
    |\mathcal{A}_i|& + |\mathcal{A}_j| + \sum_{\ell\in \mathcal{I}_i\setminus \{k\}}|\mathcal{B}_{\ell,i}|+ \sum_{\ell\in \mathcal{I}_j\setminus \{k\}}|\mathcal{B}_{\ell,j}|\notag \\
    &= \left(|\mathcal{A}_i^{\mathbf{K}_N}|- \binom{\delta_i}{t}\right) + \left(|\mathcal{A}_j^{\mathbf{K}_N}|- \binom{\delta_j}{t}\right)+ \sum_{\ell\in \mathcal{I}_i\setminus \{k\}}\binom{N-3}{t-2}+ \sum_{\ell\in \mathcal{I}_j\setminus \{k\}}\binom{N-3}{t-2}\\
    &= \left(|\mathcal{A}_i^{\mathbf{K}_N}|+|\mathcal{A}_j^{\mathbf{K}_N}|+ \sum_{\ell\in \mathcal{I}_i^{\mathbf{K}_N}\setminus \{k\}}\binom{N-3}{t-2}+ \sum_{\ell\in \mathcal{I}_j^{\mathbf{K}_N}\setminus \{k\}}\binom{N-3}{t-2}\right)\notag\\
    &\quad-\left(\binom{\delta_i}{t}+\binom{\delta_j}{t}+\delta_i\binom{N-3}{t-2}+\delta_j\binom{N-3}{t-2}\right)\\
    &=D^{\mathbf{K}_N}- \Delta_k.
\end{align}
The resulting rate is
\begin{align}
    R(G)& = \frac{KL}{\sum_{k=1}^K(D^{\mathbf{K}_N}-\Delta_k)}\\
    &= \frac{1}{\frac{D^{\mathbf{K}_N}}{L}-\frac{\sum_{k=1}^K\Delta_k}{KL}},
\end{align}
which gives the desired rate since $R^{\Pi_t}(\mathbf{K}_N)=\frac{L}{D^{\mathbf{K}_N}}$. The maximization over all $t\in [N-1]$ yields the greatest lower bound for $C(G)$.

Privacy against server $n$ holds because, for a fixed $t$, the queries received by the server follow the same structure, for every $\theta\in \mathcal{I}_n$. That is, the query structure is such that the user requests for equal number of symbols of all messages in $\mathcal{W}_n$, where the symbol indices are hidden because of the user-private independent permutations. This is because, upon the conversion from $\mathcal{A}_n^{\mathbf{K}_N}$ to $\mathcal{A}_n$, every $\ell\in \mathcal{I}_n$ appears equal number of times across all subsets in $\mathcal{A}_n$. The multiset $\mathcal{A}_n$ consists of $\binom{N-1}{t}-\binom{\delta_n}{t}$ subsets (with repetitions), where each subset is of size $\rho\in [t]$.  For a fixed $t$, any $\rho$-subset of $\mathcal{I}_n$ is contained in exactly $\binom{\delta_n}{t-\rho}$ subsets of $\mathcal{A}_n^{\mathbf{K}_N}$.  Moreover, there are $\binom{\deg(n)-1}{\rho-1}$ unique $\rho$-subsets which contain $\ell$. Therefore, the number of $\rho$-subsets in $\mathcal{A}_n$, that $\ell\in \mathcal{I}_n$ appears in is precisely $\binom{\delta_n}{t-\rho}\binom{\deg(n)-1}{\rho-1}$, which is independent of $\ell$.

\begin{example}\label{ex:complete_based_bi}
We illustrate the scheme on $\mathbf{K}_{2,3}$, where the optimal choice of $t$ is $2$, and $L=3$. With $\theta=1$, starting from $\mathbf{K}_5$, we construct the multisets $\mathcal{A}_n$ and $\mathcal{B}_{\ell,n}$ for $n=1,3$ and $\ell \in \{2,3,4\}$, given by
\begin{align}
    \mathcal{A}_1 =& \left(\{1\},\{2\},\{3\},\{1,2\},\{1,3\},\{2,3\}\right), \quad \mathcal{A}_3 = \left(\{1\},\{1\},\{4\},\{4\},\{1,4\}\right),\\
    &\mathcal{B}_{2,1} = \{\{1,2\}\}, \quad \mathcal{B}_{3,1} = \{\{1,3\}\},  \quad \mathcal{B}_{4,2} =\{\{1,4\}\}.
\end{align}
From here, we deduce the corresponding answers as shown in Table~\ref{tab:answers_bip_2}, and the rate is $\frac{6}{14}=\frac{3}{7}$.   
\end{example}

\begin{table}[t]
    \centering
    \begin{tabular}{|c|c|c|c|c|}
    \hline
    server 1 & server 2 & server 3 & server 4 & server 5\\
    \hline
        $a_1, b_1, c_1$ & $d_3$ & $a_4,d_1$ & $b_2$ & $c_2$\\
        $a_2+b_2$ & & $a_5,d_2$ & &\\
        $a_3+c_2$ & & $a_6+d_3$ & &\\
        $b_3+c_3$ & & & &\\
        \hline
    \end{tabular}
    \caption{Retrieval scheme for $\mathbf{K}_{2,3}$ for Example~\ref{ex:complete_based_bi}.}
    \label{tab:answers_bip_2}
\end{table}

\subsection{Proof of Theorem \ref{thm:ach_bipartite}}\label{proof:bipartite}
Assume that $L=1$. Given $G$ bipartite, with parts $V_1$ and $V_2$, let 
\begin{align}
    m^* = \arg \min_{m\in \{1,2\}} \sum_{n\in V_m} \deg (n)^2.
\end{align}
For $\theta=k$, suppose $W_k$ is stored on servers $i$ and $j$. Assume without loss of generality that $i\in V_{m^*}$ and $j\notin V_{m^*}$. Then, the user queries for all messages from server $i$, downloading $\deg(i)$ messages and nothing at all from server $j$.\footnote{This resembles the scheme of single-server PIR, where to preserve privacy, one must download all the messages stored on it.} In short, the user downloads the message set $\mathcal{W}_n$ if $\theta \in \mathcal{I}_n$ and $n \in V_{m^*}$.  The resulting total download across all $\theta$ is
\begin{align}
\sum_{k=1}^K D_k &= 
\sum_{n \in V_{m^*}}\deg(n)L \left(\sum_{k=1}^K \mathds{1}\{k\in \mathcal{I}_n\}\right)\label{eq:sum_indicators}\\
&= \sum_{n\in V_{m^*}} \deg(n)^2 L,\label{eq:d_total_bipartite}
\end{align}
since the inner sum in \eqref{eq:sum_indicators} is $\deg(n)$. The ratio between $KL$ and \eqref{eq:d_total_bipartite} gives the desired rate. Local privacy follows since whenever the user downloads $\mathcal{W}_n$ from server $n\in V_{m^*}$, only the fact that $\theta\in \mathcal{I}_n$ is revealed to it.  

\begin{remark}
    Interestingly, this scheme provides standard privacy as PIR (not limited to local privacy) against any server $n\in V\setminus V_{m^*}$. This is because, irrespective of the realization of $\theta$, the user does not communicate with these servers.
\end{remark}

Continuing with the running example of $\mathbf{K}_{2,3}$, we have $V_1 = \{1,2\}$ and $V_2 = \{3,4,5\}$. Then, $m^* = 2$, and the answers when $\theta = 1$ are $A_3^{[1]}=\{W_1,W_4\}$, which results in the rate $\frac{1}{2}$.

\subsection{Proof of Theorem~\ref{thm:gen_graph_ach}}\label{proof_L=1}
For this scheme, we assume $L=1$ for all messages.\footnote{Without loss of generality, we assume each message is a single bit, since the same query can be applied to each bit individually across the entire message length for any $L>1$.} The user privately generates $K$ bits, $h_\ell, ~\ell\in [K]$, uniformly at random and independently from $\{0,1\}$. For server $n$, let $\mathcal{W}_n$ be arranged in an increasing order of the elements in $\mathcal{I}_n$ into the vector $\bm{W}_n\in \mathbb{F}_2^{\deg(n)}$, i.e., letting $\mathcal{I}_n = \{\ell_1,\ell_2,\ldots,\ell_{\deg(n)}\}$ where $\ell_1<\ell_2<\ldots<\ell_{\deg(n)}$, we write
\begin{align}
    \bm{W}_n &=
    \begin{bmatrix}
        W_{\ell_1} & W_{\ell_2} & \ldots & W_{\ell_{\deg(n)}}
    \end{bmatrix}^\top.
\end{align}
Similarly, let 
\begin{align}
    \bm{h}_n =  \begin{bmatrix}h_{\ell_1}  & h_{\ell_2} & \ldots & h_{\ell_{\deg(n)}}\end{bmatrix}^\top
\end{align}
be the binary random vector constituting the bits $\{h_\ell, ~\ell\in \mathcal{I}_n\}$, arranged in the same order. 

Let $W_k$ be stored on server $i$ and server $j$, where without loss of generality, $i<j$. To retrieve $W_{k}$, the user sends a binary query vector each, to server $i$ and server $j$, while sending nothing to the remaining servers. Assuming that $W_k$ appears in the $p$th row of $\bm{W}_i$ and in the $m$th row of $\bm{W}_j$, i.e., $\bm{W}_i(p)=\bm{W}_j(m) = W_k$, where $p\in [\deg(i)], ~m\in [\deg(j)]$, the queries sent are
\begin{align}
    Q_i^{[k]} = \bm{h}_i, \qquad Q_j^{[k]} = \bm{h}_j+\bm{e}_m,
\end{align}
where $\bm{h}_i(p)=\bm{h}_j(m) = h_k$, and ``$+$'' denotes bit-wise addition, and $\bm{e}_m$ is the $\deg(j)$ unit column vector with $1$ at row $m$. In response, the answer $A_n^{[k]}$ returned by server $n$ is a vector whose $\nu$th entry $A_n^{[k]}(\nu)$ is given by the rule
\begin{align}
    A_n^{[k]}(\nu) = \begin{cases} Q_n^{[k]}(\nu)\cdot \bm{W}_n (\nu), & Q_n^{[k]}(\nu)=1,\\
        \emptyset , & Q_n^{[k]}(\nu)= 0,
    \end{cases}
\end{align}
where ``$\cdot$'' denotes scalar multiplication in $\mathbb{F}_{2}$. If the $\nu$th query entry of server $n$ is non-zero, then $A_n^{[k]}(\nu)=\bm{W}_n(\nu)$. Otherwise, the corresponding entry in $A_n^{[k]}$ is omitted. Thus, the length of answer vectors $A_i^{[k]}$ and $A_j^{[k]}$, for a fixed realization of query-pair, is given by the number of ones in the query vectors $Q_i^{[k]}$ and $Q_j^{[k]}$, respectively. 

The download cost $D_k$ is given by the sum of answer lengths $A_i^{[k]}$ and $A_j^{[k]}$, averaged over all realizations of $\{h_\ell, ~ \ell\in \mathcal{I}_i \cup \mathcal {I}_j\}\in \{0,1\}^{\deg(i)+\deg(j)-1}$. Note that, the desired message $W_k$ is downloaded  either from server $i$ or from server $j$, which contributes $1$ bit to the download cost. Let $X_i\in [0:\deg(i)-1]$ and $X_j\in [0:\deg(j)-1]$ be the number of ones in $Q_i^{[k]}$ and $Q_j^{[k]}$, excluding the entries $h_k$ and $h_k+1$, respectively. Clearly, since $h_\ell$ is uniform on $\{0,1\}$, and independent, $X_i\sim \text{Binomial}(\deg(i)-1, 0.5)$ and $X_j \sim \text{Binomial}(\deg(j)-1, 0.5)$. Therefore, taking expectation over $\bm{h}_i$ and $\bm{h}_j$ yields
\begin{align}
    D_k &= 1+\mathbb{E}\left[X_i\right]+ \mathbb{E}\left[X_j\right]\\
    &=1+ \frac{\deg(i)-1}{2}+\frac{\deg(j)-1}{2}\\
    &= \frac{\deg(i)+\deg(j)}{2}.
\end{align}
Summing across all $k$, we get
\begin{align}
    \sum_{k=1}^K D_k
    &=\frac{1}{2}\sum_{\{i,j\}\in E} (\deg(i)+\deg(j))  \\
    &= \frac{1}{2}\cdot\sum_{i\in V} \deg(i) \left(\sum_{j\in V\setminus \{i\}} \mathds{1}\{\{i,j\}\in E\}\right)\\
    &=\frac{1}{2} \cdot \sum_{i\in V} \left(\deg(i) \cdot \deg(i)\right).
\end{align}
Finally, rate is given by $\frac{K}{\sum_{k=1}^K D_k}=\frac{2K}{\sum_{i\in V}\deg(i)^2}$. 

For each desired message index, $\{k\}=\mathcal{I}_i\cap \mathcal{I}_j$, server $n\in \{i,j\}$ receives a uniformly random binary vector of $\deg(n)$ length as a query. This is a probabilistic scheme from the equivalent viewpoint that server $n$ is sent one of the feasible query sets, depending upon the realization of $\bm{h}_n$, which are identically distributed irrespective of $\theta\in \mathcal{I}_n$.

\begin{table}[t]
    \centering
    \begin{tabular}{|c|c|c||c|c|c|}
    \hline
    $(h_1 h_2 h_3 h_4)$& server 1& server 3 & $(h_1 h_2 h_3 h_4)$ & server 1 & server 3\\
    \hline
    $(0000)$ & $\emptyset$ & $W_1$ & $(0001)$ & $\emptyset$ & $W_1,W_4$ \\
    \hline
    $(0010)$ & $W_3$ & $W_1$ & $(0011)$ & $W_3$ & $W_1,W_4$ \\
    \hline
    $(0100)$ & $W_2$ & $W_1$ & $(0101)$ & $W_2$ & $W_1,W_4$ \\
    \hline
    $(0110)$ & $W_2,W_3$ & $W_1$ & $(0111)$ & $W_2,W_3$ & $W_1,W_4$ \\
    \hline
    $(1000)$ & $W_1$ & $\emptyset$ & $(1001)$ & $W_1$ & $W_4$ \\
    \hline
    $(1010)$ & $W_1,W_3$ & $\emptyset$ & $(1011)$ & $W_1,W_3$ & $W_4$ \\
    \hline
    $(1100)$ & $W_1,W_2$ & $\emptyset$ & $(1101)$ & $W_1,W_2$ & $W_4$ \\
    \hline
    $(1110)$ & $W_1,W_2,W_3$ & $\emptyset$ & $(1111)$ & $W_1,W_2,W_3$ & $W_4$ \\
    \hline
    \end{tabular} 
    \caption{Table showing all possible queries and the downloads for $\theta=1$ for $\mathbf{K}_{2,3}$.}
    \label{tab:query_sets}
\end{table}

Continuing with the running example of $\mathbf{K}_{2,3}$, for $\theta=1$, there are $2^4 = 16$ possible queries, depending on the realizations of $h_1,h_2,h_3,h_4$, as shown in Table~\ref{tab:query_sets}. Each query is chosen with equal probability $2^{-4}$, and the resulting rate is given by $\frac{2}{5}$.

\section{Proofs of Capacity Upper Bounds}\label{sec:ubnd proofs}
In this section, we derive the results on capacity upper bounds. Consider a storage based on a connected, simple graph $G$ with $N$ vertices, and $K$ edges. For the desired index $\theta=k$, the number of downloaded symbols from all servers is,
\begin{align}
    D_k &=  \sum_{n=1}^N H(A_n^{[k]})\\
    &\geq \sum_{n=1}^N H(A_n^{[k]}|\mathcal{Q})\\
    &\geq H(A_{[N]}^{[k]}|\mathcal{Q})\\
    &=   H(W_k|A_{[N]}^{[k]},\mathcal{Q}) +H(A_{[N]}^{[k]}|\mathcal{Q}) \label{eq:upper_bnd_decodability}\\    
    &= H(W_k|\mathcal{Q})+  H(A_{[N]}^{[k]}|W_k,\mathcal{Q}) - H(A_{[N]}^{[k]}|\mathcal{W},\mathcal{Q})\label{eq:upper_bnd_answer_function}\\
    & = L+ I(\mathcal{W}\setminus \{W_k\};A_{[N]}^{[k]}|W_k,\mathcal{Q}),\label{eq:interference}
\end{align}
where \eqref{eq:upper_bnd_decodability} follows from decodability \eqref{r_4}, and the last term in \eqref{eq:upper_bnd_answer_function} is zero due to \eqref{r_3}. The second term in \eqref{eq:interference} is the information about the non-desired messages $\mathcal{W}\setminus W_k$, i.e., interference contained in the answers $A_{[N]}^{[k]}$. Then, the lower bound on the total download  $\sum_{k=1}^K D_k$ is $KL$ added to the $K$ interference terms. To obtain suitable bounds on these quantities, we require the following lemmas that any local PIR scheme must satisfy. The first lemma is a modification of \cite[Proposition 2]{SGT23}, in the sense that, for each $n$, the support set of $\theta$ is precisely $\mathcal{I}_n$, instead of $[K]$ for each $n$.
\begin{lemma}\label{lem:user_priv}
For any $\mathcal{J}\subset [K]$, define the message set $W_{\mathcal{J}} = \{W_{\ell}:\ell\in \mathcal{J}\}$. Then, by the local user privacy against server $n$, we have
\begin{align}
    I(\theta; A_n^{[\theta]}|Q_n^{[\theta]}, W_{\mathcal{J}}) = 0, \quad \theta \in \mathcal{I}_n, ~n\in [N].
\end{align}
\end{lemma}
\begin{Proof}
This is equivalent to showing that for any $k,k'\in \mathcal{I}_n$ with $k\neq k'$
\begin{align}
    H(A_n^{[k]}|Q_n^{[k]},W_\mathcal{J})=H(A_n^{[k']}|Q_n^{[k']},W_\mathcal{J}).
\end{align}
Let $\mathcal{W}_n' = \mathcal{W}_n \cap W_{\mathcal{J}}$. Note that,
\begin{align}
    H(A_n^{[k]}|Q_n^{[k]},W_\mathcal{J})&=   H(A_n^{[k]}|Q_n^{[k]},\mathcal{W}_n')-I(W_{\mathcal{J}}\setminus \mathcal{W}_n';A_n^{[k]}|Q_n^{[k]},\mathcal{W}_n')\\
    &=H(A_n^{[k']}|Q_n^{[k']},\mathcal{W}_n')-I(W_{\mathcal{J}}\setminus \mathcal{W}_n';A_n^{[k]}|Q_n^{[k]},\mathcal{W}_n'),\label{eq:ans indep of non intersecting msgs}
\end{align}
where \eqref{eq:ans indep of non intersecting msgs} follows from local user privacy requirement \eqref{r_2} for any $k,k'\in \mathcal{I}_n$. It remains to show that $I(W_{\mathcal{J}}\setminus \mathcal{W}_n';A_n^{[k]}|Q_n^{[k]},\mathcal{W}_n')=0$, which is true since,
\begin{align}
    I(W_{\mathcal{J}}\setminus \mathcal{W}_n';A_n^{[k]}|Q_n^{[k]},\mathcal{W}_n')
    &\leq  I(W_{\mathcal{J}}\setminus \mathcal{W}_n';A_n^{[k]}, \mathcal{W}_n\setminus \mathcal{W}_n'|Q_n^{[k]},\mathcal{W}_n')\\
    &= I(W_{\mathcal{J}}\setminus \mathcal{W}_n';\mathcal{W}_n\setminus \mathcal{W}_n'|Q_n^{[k]},\mathcal{W}_n')+I(W_{\mathcal{J}}\setminus \mathcal{W}_n';A_n^{[k]}|Q_n^{[k]},\mathcal{W}_n)\\
    &\leq I(W_{\mathcal{J}}\setminus \mathcal{W}_n';\mathcal{W}_n|Q_n^{[k]})+ H(A_n^{[k]}|Q_n^{[k]},\mathcal{W}_n) \label{eq:both terms zero}\\
   &=0.
\end{align}
The first term in \eqref{eq:both terms zero} is $0$ by the independence of the disjoint message sets, and their independence from the query, while the second term is $0$ due to \eqref{r_3}. 
\end{Proof}

The next lemma states that given any subset of messages, the answer from server $n$ depends solely on the query sent to server $n$, and is independent of the rest of $\mathcal{Q}$. Note that, the lemma applies to the standard PIR setting, irrespective of the servers' storage.

\begin{lemma}\label{lem:answer_indep_randomness_given_query}
    For any subset $\mathcal{J}\subset [K]$, and any $k\in [K]$
    \begin{align}
        H(A_n^{[k]}|W_{\mathcal{J}}, Q_n^{[k]})&=H(A_n^{[k]}|W_{\mathcal{J}},Q_n^{[k]},\mathcal{Q})\label{eq:a}\\
        &=H(A_n^{[k]}|W_{\mathcal{J}},\mathcal{Q})\label{eq:b}.
    \end{align}
\end{lemma}

\begin{Proof}
For \eqref{eq:a}, we show that $I(A_n^{[k]};\mathcal{Q}|W_{\mathcal{J}}, Q_n^{[k]})=0$. Denoting $\mathcal{W}_n'$ as $\mathcal{W}_n \cap W_{\mathcal{J}}$, we have
\begin{align}
    I(A_n^{[k]};\mathcal{Q}|W_{\mathcal{J}}, Q_n^{[k]}) 
    \leq& I(A_n^{[k]},\mathcal{W}_{n}\setminus\mathcal{W}_n';\mathcal{Q}|W_{\mathcal{J}}, Q_n^{[k]})\\
    =& I(\mathcal{W}_{n}\setminus\mathcal{W}_n';\mathcal{Q}|W_{\mathcal{J}}, Q_n^{[k]})\!+\!I(A_n^{[k]};\mathcal{Q}|\mathcal{W}_n, W_{\mathcal{J}}\setminus \mathcal{W}_n',Q_n^{[k]})\label{eq:last term is zero}\\
    \leq&I(\mathcal{W}_{n}\setminus\mathcal{W}_n';\mathcal{Q}|W_{\mathcal{J}}, Q_n^{[k]}) + I(W_{\mathcal{J}};\mathcal{Q}|Q_n^{[k]})\label{eq:follows by answer generation}\\
    =&I(\mathcal{W}_n, W_{\mathcal{J}}\setminus \mathcal{W}_n';\mathcal{Q}|Q_n^{[k]})\\
    \leq& I(\mathcal{W}_n, W_{\mathcal{J}}\setminus \mathcal{W}_n';\mathcal{Q},Q_n^{[k]})=0,
\end{align}
where \eqref{eq:follows by answer generation} follows since the second term of \eqref{eq:last term is zero} is zero since $H(A_n^{[k]}|\mathcal{W}_n,W_{\mathcal{J}}\setminus \mathcal{W}_n',Q_n^{[k]})=0$, and the last equality follows by \eqref{r_1}. The equality \eqref{eq:b} holds since $H(Q_n^{[k]}|\mathcal{Q}) = 0$. 
\end{Proof}

The next lemma is a re-statement of \cite[Lemma 3]{SGT23}, and applies to our local PIR setting as well, by the decodability requirement \eqref{r_4}. 

\begin{lemma}[Lemma 3 in \cite{SGT23}]\label{lem:graph_decodability_lemma}
    For any two servers $i,j$ with $\{i,j\}\in E$, and for any $k\in \mathcal{I}_i \cap \mathcal{I}_j$, a local PIR scheme on the graph $G$,  satisfies
    \begin{align}
        H(A_i^{[k]}|\mathcal{W}\setminus \{W_k\}, \mathcal{Q})+H(A_j^{[k]}|\mathcal{W}\setminus \{W_k\}, \mathcal{Q})\geq L.
    \end{align}
\end{lemma}

\subsection{Proof of Theorem~\ref{thm:cyclic_converse}}\label{proof:cyclic_converse}
Given $\mathbf{C}_N$, let the storage of server $n$ be $\mathcal{W}_n = \{W_n, W_{n-1}\}$, with the convention that $W_0=W_N$. Throughout the proof, we consider the message and server indices modulo $N$. For $\theta=k$, the lower bound on the interference term is given by
\begin{align}
    I(\mathcal{W}\setminus \{W_k\};A_{[N]}^{[k]}|W_k,\mathcal{Q})&\geq I(W_{k-1}, W_{k+1} ; A_{[N]}^{[k]}|W_k, \mathcal{Q})\\
    & = I(W_{k-1};A_{[N]}^{[k]}|W_k,\mathcal{Q})+ I(W_{k+1};A_{[N]}^{[k]}|W_k, W_{k-1},\mathcal{Q})\\
    &\geq I(W_{k-1};A_{k}^{[k]}|W_k,\mathcal{Q})+ I(W_{k+1};A_{k+1}^{[k]}|W_k, W_{k-1},\mathcal{Q})\\
    &=  I(W_{k-1};A_{k}^{[k-1]}|W_k,\mathcal{Q})+ I(W_{k+1};A_{k+1}^{[k+1]}|W_k, W_{k-1},\mathcal{Q})\label{eq:change index}\\
    &=H(A_k^{[k-1]}|W_k,\mathcal{Q}) + H(A_{k+1}^{[k+1]}|W_k,  W_{k-1},\mathcal{Q})\label{eq:ans_deterministic}\\
    &\geq H(A_k^{[k-1]}|\mathcal{W}\setminus \{W_{k-1}\},\mathcal{Q}) + H(A_{k+1}^{[k+1]}|\mathcal{W}\setminus \{W_{k+1}\},\mathcal{Q}),
\end{align}
where \eqref{eq:change index} follows by applying Lemma~\ref{lem:user_priv} and Lemma~\ref{lem:answer_indep_randomness_given_query}, \eqref{eq:ans_deterministic} follows since $H(A_k^{[k-1]}|W_k,W_{k-1},\mathcal{Q})=0$ and $H(A_{k+1}^{[k+1]}|W_{k+1},W_k,  W_{k-1},\mathcal{Q})=0$. Now, summing $D_k-L$ over all $k\in [N]$, we have
\begin{align}
    \sum_{k=1}^N D_k - NL \geq & \sum_{k=1}^N \left( H(A_k^{[k-1]}|\mathcal{W}\setminus \{W_{k-1}\},\mathcal{Q}) + H(A_{k+1}^{[k+1]}|\mathcal{W}\setminus \{W_{k+1}\},\mathcal{Q}) \right)\\
    &= H(A_1^{[N]}|\mathcal{W}\setminus \{W_N\}, \mathcal{Q})+H(A_2^{[2]}|\mathcal{W}\setminus \{W_2\},\mathcal{Q})\notag\\
    & \quad+H(A_2^{[1]}|\mathcal{W}\setminus \{W_1\}, \mathcal{Q}) + H(A_3^{[3]}|\mathcal{W}\setminus \{W_3\},\mathcal{Q})\notag\\
    &\quad+ H(A_3^{[2]}|\mathcal{W}\setminus \{W_2\}, \mathcal{Q}) + H(A_4^{[4]}\mathcal{W}\setminus \{W_4\},\mathcal{Q})\notag \\
    &\quad + \ldots+ H(A_{N}^{[N-1]}|\mathcal{W}\setminus \{W_{N-1}\},\mathcal{Q})+H(A_1^{[1]}|\mathcal{W}\setminus \{W_1\}, \mathcal{Q})\\
    &=\sum_{k=1}^N H(A_k^{[k]}|\mathcal{W}\setminus\{W_k\},\mathcal{Q})+ H(A_{k+1}^{[k]}|\mathcal{W}\setminus\{W_k\},\mathcal{Q})\label{eq:grouping_answers}\\
    &= NL, \label{eq:msg_decodable}
\end{align}
where \eqref{eq:grouping_answers} follows by grouping the answer pairs with the same message index (superscript), and \eqref{eq:msg_decodable} holds by applying Lemma~\ref{lem:graph_decodability_lemma}. 

\subsection{Proof of Theorem~\ref{thm:path_converse}}\label{proof:path_converse}
Given $\mathbf{P}_N$, let the storage of server $n$ be 
\begin{align}
\mathcal{W}_n &=
 \begin{cases}
      \{W_1\}, &n=1,\\
      \{W_n, W_{n-1}\}, & n\in [2:N-1],\\
    \{W_{N-1}\}, & n=N.
 \end{cases} 
\end{align} 
Our goal is to lower bound the second term in \eqref{eq:interference} by following arguments similar to those in Section~\ref{proof:cyclic_converse}. First, for $k=1$,
\begin{align}
   I(\mathcal{W}\setminus \{W_1\};A_{[N]}^{[1]}|W_1,\mathcal{Q})&\geq I(W_{2} ; A_{[N]}^{[1]}|W_1, \mathcal{Q})\\
   &\geq I(W_{2} ; A_2^{[1]}|W_1, \mathcal{Q})\\
   &=H(A_2^{[1]}|W_1, \mathcal{Q})\\
   &= H(A_2^{[2]}|W_1, \mathcal{Q})\\
   &\geq H(A_2^{[2]}|\mathcal{W}\setminus \{W_2\}, \mathcal{Q}).
\end{align}
Similarly, for $k=N-1$, it follows by symmetry that
\begin{align}
   I(\mathcal{W}\setminus \{W_{N-1}\};A_{[N]}^{[N-1]}|W_{N-1},\mathcal{Q})&\geq 
   H(A_{N-1}^{[N-2]}|\mathcal{W}\setminus \{W_{N-2}\}, \mathcal{Q}).
\end{align}
For $k\in [2:N-2]$, we obtain
\begin{align}
    I(\mathcal{W}\setminus \{W_k\};A_{[N]}^{[k]}|W_k,\mathcal{Q})&\geq I(W_{k-1}, W_{k+1} ; A_{[N]}^{[k]}|W_k, \mathcal{Q})\\
    & = I(W_{k-1};A_{[N]}^{[k]}|W_k,\mathcal{Q})+ I(W_{k+1};A_{[N]}^{[k]}|W_k, W_{k-1},\mathcal{Q})\\
    &\geq I(W_{k-1};A_{k}^{[k]}|W_k,\mathcal{Q})+ I(W_{k+1};A_{k+1}^{[k]}|W_k, W_{k-1},\mathcal{Q})\\
    &=  I(W_{k-1};A_{k}^{[k-1]}|W_k,\mathcal{Q})+ I(W_{k+1};A_{k+1}^{[k+1]}|W_k, W_{k-1},\mathcal{Q})\\
    &=H(A_k^{[k-1]}|W_k,\mathcal{Q}) + H(A_{k+1}^{[k+1]}|W_k,  W_{k-1},\mathcal{Q})\\
    &\geq H(A_k^{[k-1]}|\mathcal{W}\setminus \{W_{k-1}\},\mathcal{Q}) + H(A_{k+1}^{[k+1]}|\mathcal{W}\setminus \{W_{k+1}\},\mathcal{Q}).
\end{align}
Now, summing $D_k-L$ over all $k\in [N-1]$, we have
\begin{align}
    \sum_{k=1}^{N-1} D_k - (N-1)L &\geq  H(A_2^{[2]}|\mathcal{W}\setminus \{W_{2}\},\mathcal{Q})\notag\\
    &\quad+\sum_{k=2}^{N-2} \left( H(A_k^{[k-1]}|\mathcal{W}\setminus \{W_{k-1}\},\mathcal{Q}) + H(A_{k+1}^{[k+1]}|\mathcal{W}\setminus \{W_{k+1}\},\mathcal{Q}) \right)\notag\\
    &\quad+H(A_{N-1}^{[N-2]}|\mathcal{W}\setminus \{W_{N-2}\},\mathcal{Q})\\
    &= H(A_2^{[2]}|\mathcal{W}\setminus \{W_{2}\},\mathcal{Q})+H(A_2^{[1]}|\mathcal{W}\setminus \{W_{1}\}, \mathcal{Q}) + H(A_3^{[3]}|\mathcal{W}\setminus \{W_{3}\},\mathcal{Q})\notag\\
    &\quad+ H(A_3^{[2]}|\mathcal{W}\setminus \{W_{2}\}, \mathcal{Q}) + H(A_4^{[4]}|\mathcal{W}\setminus \{W_{4}\},\mathcal{Q})+\notag \\
    &\quad  \ldots+ H(A_{N-2}^{[N-3]}|\mathcal{W}\setminus \{W_{N-3}\},\mathcal{Q})+H(A_{N-1}^{[N-1]}|\mathcal{W}\setminus \{W_{N-1}\}, \mathcal{Q}) \notag\\
    &\quad+H(A_{N-1}^{[N-2]}|\mathcal{W}\setminus \{W_{N-2}\},\mathcal{Q})\label{eq:re-writing}\\
    &\geq \sum_{k=2}^{N-2} \left(H(A_k^{[k]}|\mathcal{W}\setminus \{W_k\},\mathcal{Q})+H(A_{k+1}^{[k]}|\mathcal{W}\setminus \{W_k\},\mathcal{Q}\right)\label{eq:follows from dropping the terms}\\
    &\geq (N-3)L, \label{eq:follows from lem3} 
\end{align}
where \eqref{eq:follows from dropping the terms} follows from \eqref{eq:re-writing} by dropping the entropy terms $H(A_2^{[1]}|\mathcal{W}\setminus \{W_1\},\mathcal{Q})$ and $H(A_{N-1}^{[N-1]}|\mathcal{W}\setminus \{W_{N-1}\},\mathcal{Q})$, and \eqref{eq:follows from lem3} follows from Lemma~\ref{lem:graph_decodability_lemma}.
\subsection{Proof of Remark~\ref{rmk:half_upper_bnd}}\label{pf:half_upper_bnd}
We consider the PIR systems for which $\deg(n) = |\mathcal{I}_n|\geq 2$ for all $n\in [N]$. Let each index set $\mathcal{I}_n=\{\ell_t, ~t\in \deg(n)\}$ be arranged into the vector,
\begin{align}
    \bm{I}_n = \begin{bmatrix}
        \ell_1 & \ell_2 &\ldots & \ell_{\deg(n)}
    \end{bmatrix},
\end{align}such that $\ell_1 < \ell_2 < \ldots < \ell_{\deg(n)}$. Also, consider the cyclic order of the indices in server $n$ given by $\ell_1\to \ell_2 \to \ldots\to \ell_{\deg(n)}\to \ell_1$. Now, let servers $i$ and $j$ share the message $W_k$, and let $k=\bm{I}_i(m_i)=\bm{I}_j(m_j)$, where $m_i\in [\deg(i)]$ and $m_j \in [\deg(j)]$. Further, let $k_1=\bm{I}_i(m_i+1)$ and $k_2=\bm{I}_j(m_j+1)$, where the indices are taken in a cyclic order. Clearly, since the degrees of all vertices is at least $2$, $k_1$ and $k_2$ are distinct from $k$, and $k_1\neq k_2$ since $G$ is simple. Then, the following lower bound holds.
\begin{align}
    D_k -L\geq&  I(\mathcal{W}\setminus \{W_k\};A_{[N]}^{[k]}|W_k,\mathcal{Q}) \\
    \geq& I(\mathcal{W}_i \cup \mathcal{W}_j \setminus \{W_k\}; A_{[N]}^{[k]}|W_k,\mathcal{Q})\\
    = & I(\mathcal{W}_i \setminus \{W_k\}; A_{[N]}^{[k]}|W_k,\mathcal{Q})+ I( \mathcal{W}_j \setminus \{W_k\}; A_{[N]}^{[k]}|\mathcal{W}_i \setminus \{W_k\},W_k,\mathcal{Q})\\ 
    \geq&  I(\mathcal{W}_i\setminus\{W_k\};A_i^{[k]}|W_k,\mathcal{Q})+I(\mathcal{W}_j\setminus\{W_k\};A_j^{[k]}|\mathcal{W}_i\setminus\{W_k\},W_k,\mathcal{Q})\\
    =&  H(A_i^{[k]}|W_k,\mathcal{Q})- H(A_i^{[k]}|\mathcal{W}_i,\mathcal{Q})+H(A_j^{[k]}|\mathcal{W}_i,\mathcal{Q}) - H(A_j^{[k]}|\mathcal{W}_i,\mathcal{W}_j,\mathcal{Q})\\
     =&  H(A_i^{[k_1]}|W_k,\mathcal{Q})+H(A_j^{[k_2]}|\mathcal{W}_i,\mathcal{Q}) \\
    \geq& H(A_i^{[k_1]}|\mathcal{W}\setminus \{W_{k_1}\},\mathcal{Q})+H(A_j^{[k_2]}|\mathcal{W}\setminus \{W_{k_2}\},\mathcal{Q}).
\end{align}
Let $\{k_1\}=\mathcal{I}_i\cap \mathcal{I}_{n_1}$, and $\{k_2\}=\mathcal{I}_j\cap \mathcal{I}_{n_2}$ for some $n_1,n_2\in [N],~ n_1, n_2\neq i,j$. In particular, let $k_1=\bm{I}_{n_1}(m_{n_1})$ and $k_2=\bm{I}_{n_2}(m_{n_2})$, where $m_{n_1}\in[\deg(n_1)]$ and $m_{n_2}\in[\deg(n_2)]$. Notice that, for $k' = \bm{I}_{n_1}(m_{n_1}-1)$, it holds that $H(A_{n_1}^{[k_1]}|\mathcal{W}\setminus \{W_{k_1}\},\mathcal{Q})$ is one of the terms on the right hand side of the inequality bounding $D_{k'}-L$. Similarly, for $k'' = \bm{I}_{n_2}(m_{n_2}-1)$, it holds that $H(A_{n_2}^{[k_2]}|\mathcal{W}\setminus \{W_{k_2}\},\mathcal{Q})$ is one of the terms on the right hand side of the inequality bounding $D_{k''}-L$. Such $k'$ and $k''$ exist because the degrees of $n_1$ and $n_2$ are at least $2$. Summing the inequality over all $k$ yields $KL$ on the right hand side, by applying Lemma~\ref{lem:graph_decodability_lemma} to each pair of answers with the same index.  

\section{Generalizing to Multigraphs}\label{sec:multigraphs}
In this section, we extend the local PIR results from simple graphs to $r$-multigraphs, where $r\geq 1$. We begin with the problem setting and introduce the new notations associated with this extension. Next, we state the results in Section~\ref{multigraph_results}, followed by their proofs in Section~\ref{multigraph_proofs}. 

\subsection{Problem Setting}
In the simple graph $G=(V,E)$, let $W_k, ~k\in [K']$ be the unique message stored at servers $i,j\in [N]$, where $|E|=\frac{K}{r} = K'$. Then, the $r$-multigraph version of $G$ is represented as  $G^{(r)}=(V,E^{(r)})$, where $E^{(r)}$ denotes the set of edges indexed as $\{(\ell,1),(\ell,2),\ldots,(\ell,r), ~\ell\in [K']\}$. Accordingly, for the multigraph $G^{(r)}$-based setting, let the set $\mathcal{W}_{i,j} = \mathcal{W}_i\cap \mathcal{W}_j=\{W_{k,1},W_{k,2},\ldots, W_{k,r}\}$ denote the set of $r$ unique messages stored at both servers $i$ and $j$. Let $\theta = (s,\tau)$, where $s\in [K']$ and $\tau\in [r]$, be the desired message index which is picked uniformly at random from $[K']\times [r]$.

The index set $\mathcal{I}_n$, here, is the set of $r\deg(n)$ tuples $(\ell,\tau)$, such that $W_{\ell,\tau}\in \mathcal{W}_n$ for all $\tau\in [r]$. Under this setting, the local PIR problem on $G^{(r)}$ requires satisfying the information-theoretic constraints in \eqref{r_1},\eqref{r_2}-\eqref{r_4}. The rate $R(G^{(r)})$ and capacity $C(G^{(r)})$ of the multigraph-based local PIR setting follow the same definitions as the simple graph case, i.e., when $r=1$. 

\subsection{Results}\label{multigraph_results}

\begin{table}[t]
      \centering
      \renewcommand{\arraystretch}{1.35}
      \setlength{\tabcolsep}{2pt}
      \begin{tabular}{|p{4cm}|c|c||p{6cm}|}
       \hline
       \textbf{Multigraph type} & \textbf{Thm.~\ref{thm:srp_multigraph_ach}} & \textbf{Thm.~\ref{thm:gen_multigraph_ach}}  & $R_{PIR}$\\
       \hline
       Path $\mathbf{P}_N^{(r)}$ & $\frac{N-1}{2N-3}(2-2^{1-r})^{-1}$ & $\frac{N-1}{2N-3}(2-2^{1-r})^{-1}$ & $\frac{2}{N}(2-2^{1-r})^{-1}$ \\
       \hline
       Cyclic $\mathbf{C}_N^{(r)}$ & $\frac{1}{2}(2-2^{1-r})^{-1}$ & $\frac{1}{2}(2-2^{1-r})^{-1}$ & $(1-2^{-r})^{-1}\max\left\{\frac{1}{N+1},\frac{1}{N(1+2^{-r})}\right\}$ \\
       \hline
       Complete $\mathbf{K}_N^{(r)}$ & $\frac{1}{2\sqrt{N-1}}(2-2^{1-r})^{-1}$
       & $\frac{1}{N-1}(2-2^{1-r})^{-1}$& $\frac{1}{N-2^{1-r}}$ \\
       \hline
       Star $\mathbf{S}_N^{(r)}$ & $\frac{1}{\sqrt{N-1}}(2-2^{1-r})^{-1}$ & $\frac{2}{N}(2-2^{1-r})^{-1}$& $\frac{2}{N}(2-2^{1-r})^{-1}$ \\
       \hline
       Complete, balanced bipartite $\mathbf{K}_{\frac{N}{2},\frac{N}{2}}^{(r)}$ & $\frac{1}{\sqrt{2N}}(2-2^{1-r})^{-1}$ & $\frac{2}{N}(2-2^{1-r})^{-1}$ & $\frac{1}{N\left(1-2^{-r\frac{N}{2}}\right)}$ \\
       
       \hline
    \end{tabular}
    \caption{Achievable local PIR rates and best-achievable PIR rates for common multigraph families.}
    \label{tab:multi_all_rates}
\end{table}

The achievable local rates are summarized in Table~\ref{tab:multi_all_rates}, and compared against the corresponding best-known PIR rates, which for $\mathbf{P}_N^{(r)}$ and $\mathbf{S}_N^{(r)}$ are from \cite{our_journal2025}, for $\mathbf{C}_N^{(r)}$ are from \cite{our_journal2025} and \cite{gePIR}, respectively, for $\mathbf{K}_N^{(r)}$ is from \cite{krishnan_graph} and for $\mathbf{K}_{\frac{N}{2},\frac{N}{2}}^{(r)}$ is from  \cite{gePIR}. It is worth noting that, extending the bipartite graphs scheme (Section~\ref{proof:bipartite}) directly, causes the local PIR rate of $G^{(r)}$ to decrease linearly as $\frac{1}{r}$. Thus, the rate vanishes to $0$ as $r\to \infty$, and the scheme is far from optimal. Fortunately, the other proposed schemes on simple graphs extend well to the respective multigraphs $G^{(r)}$, yielding capacity lower bounds that are bounded above zero. As established in the following, the local PIR capacity is a decreasing function of $r$, similar to its PIR counterpart. The first result connects the local PIR rate of graphs to that of $r$-multigraphs, provided that the local PIR scheme satisfies the symmetric retrieval property (SRP) \cite{our_journal2025} as defined next. 

\begin{definition}[Symmetric Retrieval Property (SRP)\cite{our_journal2025}] \label{def:srp}
    A graph-based PIR scheme is said to satisfy the symmetric retrieval property if, for any $k$, the number of symbols of $W_k$ retrieved from each of the servers storing $W_k$, is equal. Equivalently, $H(A_{i}^{[k]}|Q_{i}^{[k]},\mathcal{W}_{i}\setminus\{W_k\})=H(A_{j}^{[k]}|Q_{j}^{[k]},\mathcal{W}_{j}\setminus\{W_k\})=\frac{H(W_k)}{2}$, if $W_k$ is replicated on servers $i$ and $j$.
\end{definition}

\begin{remark}
    It can be verified that, by Theorem~\ref{thm:gen_ach_complete_based}, one obtains a local PIR scheme that satisfies SRP, for any graph $G$. Also, for edge-transitive graphs, that are regular, i.e., $\deg(i)=\deg(j)=d$, Theorem~\ref{thm:edge_trans} yields a local PIR scheme that satisfies SRP.
\end{remark}

We apply this definition and claim that there exists a local PIR scheme on any multigraph establishing a lower bound on $C\left(G^{(r)}\right)$, using the construction approach of~\cite[Section~IV.A.]{our_journal2025}.

\begin{theorem}\label{thm:srp_multigraph_ach}
    A local PIR scheme on a simple graph $G$ satisfying the SRP condition, resulting in rate $R(G)$ can be adapted to obtain a local PIR scheme on its multigraph version, $G^{(r)}$. The resulting capacity lower bound is given by
    \begin{align}
        C(G^{(r)})\geq R(G) \left(2-2^{1-r}\right)^{-1}.\label{multi_ach_srp}
    \end{align}
\end{theorem}

The scheme achieving the rate in \eqref{multi_ach_srp} requires the subpacketization level $L$ to grow by a factor of $2^{r-1}$, which may be impractical for smaller message sizes. In the next theorem, we propose another capacity lower bound for general multigraphs with minimal subpacketization, i.e., $L=1$.

\begin{theorem}\label{thm:gen_multigraph_ach}
    For the $r$-multigraph version $G^{(r)}$ of graph $G=(V,E)$ with $|V|=N$ and $|E|=\frac{K}{r}$, the capacity of local PIR is lower-bounded as
    \begin{align}\label{L=1_lowerbound_gen_multigraph}
        C(G^{(r)})\geq K\cdot \left(r(1-2^{-r})\sum_{n\in V } \deg(n)^2\right)^{-1}.
    \end{align}    
\end{theorem}

\begin{remark}
    In relation to Theorem~\ref{thm:gen_graph_ach}, if the corresponding achievable rate for $G$ is given by $R(G)$, the lower bound \eqref{L=1_lowerbound_gen_multigraph} also satisfies $C(G^{(r)})\geq \frac{K'}{(1-2^{-r})\sum_{n\in [N]}\deg(n)^2}=R(G)(2-2^{1-r})^{-1}$.
\end{remark}

For the multigraph version of graphs in Table~\ref{tab:all_rates}, the local PIR rates in the second, third and fifth columns multiplied by $(2-2^{1-r})^{-1}$ are all achievable, yielding lower bounds on $C(G^{(r)})$. Further, the capacity $C(G^{(r)})$ cannot increase as $r$ increases, because a higher $r$ implies stricter local privacy constraint. This implies that $C(G)$ is a trivial upper bound for $C(G^{(r)})$. Next, we present the upper and lower capacity bounds for the cyclic (Theorem~\ref{thm:multicycle_bounds}) and path (Theorem~\ref{thm:multipath_bounds}) multigraphs. Although there is a gap between the bounds, they improve upon the trivial upper bound for $r>1$.

\begin{theorem}[Capacity bounds for $\mathbf{C}_N^{(r)}$]\label{thm:multicycle_bounds}
    For the $r$-multigraph version of cyclic graphs, $\mathbf{C}_N^{(r)}$, with $r>1$, the local PIR capacity is  bounded as
    \begin{align}\label{eq:multicycle_bounds}
        \frac{1}{2(2-2^{1-r})}\leq C(\mathbf{C}_N^{(r)})\leq \frac{1}{3-2^{1-r}}.
    \end{align}
\end{theorem}

\begin{remark}
    Note that, the lower and upper bounds exactly match for $r=1$, while the optimality gap for $r>1$, expressed as the ratio of upper and lower bounds, settles at $\frac{4}{3}$ as $r\to \infty$. Moreover, the difference between the upper and lower bounds increases upto the maximum value of $\frac{1}{12}=0.083$ at asymptotic $r$.
\end{remark}

\begin{remark}
    In contrast, the PIR capacity for $\mathbf{C}_N^{(r)}$ is bounded as 
    \begin{align}
        \max\left\{\frac{2}{(N+1)(2-2^{1-r})},\frac{1}{N(1-2^{-2r})}\right\}  \leq C_{PIR}(\mathbf{C}_N^{(r)})\leq  \frac{1}{N-(N-1)2^{-r}},\label{eq:multicycle_PIR_bounds}
    \end{align}
    by \cite{our_journal2025} and \cite{gePIR}, respectively, where the lower bound in \eqref{eq:multicycle_bounds} exceeds the upper bound in \eqref{eq:multicycle_PIR_bounds} for $N\geq 4$.
\end{remark}

\begin{theorem}[Capacity bounds for $\mathbf{P}_N^{(r)}$]\label{thm:multipath_bounds}
    For the $r$-multigraph version of path graphs, $\mathbf{P}_N^{(r)}$, with $r>1$, the local PIR capacity is bounded as
    \begin{align}
       \frac{N-1}{(2N-3)(2-2^{1-r})} \leq C(\mathbf{P}_N^{(r)})\leq \frac{N-1}{(N-3)(3-2^{1-r})+2}.\label{multipath_N}
    \end{align}
\end{theorem}

\begin{remark}
    As $r\to \infty$, the optimality gap is $\frac{2(2N-3)}{3(N-3)}$ which is approximately $\frac{4}{3}$ for asymptotically large $N$.
\end{remark}

\begin{remark}
    The PIR capacity for $\mathbf{P}_N^{(r)}$ \cite{our_journal2025} is known exactly for even $N$  as
    \begin{align}
        C_{PIR}(\mathbf{P}_N^{(r)}) =\frac{2}{N(2-2^{1-r})}, \label{multipath_N_even}
    \end{align}
    while for odd $N$ it is bounded as
    \begin{align}
    \frac{2}{N(2-2^{1-r})} \leq C_{PIR}(\mathbf{P}_N^{(r)})\leq  \frac{2}{(N-1)(2-2^{1-r})}.\label{multipath_N_odd}
    \end{align}
    The lower bound in \eqref{multipath_N} is strictly greater for $N\geq 4$, and equal for $N=3$. 
\end{remark}

\subsection{Proofs}\label{multigraph_proofs}

\subsubsection{Proof of Theorem~\ref{thm:srp_multigraph_ach}} 
Let the local PIR scheme $\Pi$ on $G$, satisfying SRP, require the message subpacketization to be $L'$ (which is even), and the total download cost to be $D'=\sum_{k=1}^{K'} D_k$. We apply the $r$-stage algorithm proposed in \cite{our_journal2025} to such a scheme, and show that it yields a valid local PIR scheme on $G^{(r)}$ as described next. The user privately permutes the $L=2^{r-1}L'$ symbols of every message independently and uniformly at random, with $W_{\ell,\tau}(m)$ representing the $m$th permuted symbol of $W_{\ell,\tau}~, \ell\in [K'], \tau\in [r]$. For each server $n$, we partition the index set $\mathcal{I}_n$ into disjoint subsets $\mathcal{I}_{n,1},\ldots, \mathcal{I}_{n,r}$, where for each $\tau\in [r]$,
\begin{align}
    \mathcal{I}_{n,\tau} = \{(\ell,\tau):W_{\ell,\tau}\in \mathcal{W}_n\}.
\end{align}
Without loss of generality, assume that the desired message index is $\theta = (k,1)$, where $W_{k,1}\in \mathcal{W}_{i,j}$, with $i<j$. Throughout the scheme, the user communicates only with the servers in $V_k$ defined in \eqref{def_v_k} and queries for the message symbols of $\mathcal{W}_i \cup \mathcal{W}_j$. For the purpose of scheme description, we also consider the index sets $E_{k,1}, \ldots, E_{k,r}$, where for each $\tau\in [r]$,
\begin{align}
    E_{k,\tau}=\{(\ell,\tau)\in \mathcal{I}_{i,\tau}\cup \mathcal{I}_{j,\tau}\}\subseteq E^{(r)},
\end{align} 
which is the $\tau$th set of edges incident with the nodes in $V_k$. The scheme proceeds in $r$ stages as described next.

In the first stage, we apply the scheme $\Pi$ to the messages corresponding to $E_{k,1}$ to obtain the desired message symbols $W_{k,1}[1:L']$ and to $E_{k,2}, \ldots, E_{k,r}$ to retrieve the interference message symbols $\{W_{k,\tau}[1:L'], ~\tau\in [2:r]\}$. That is, $\Pi$ is applied $r$ times to disjoint message subsets. Further, assume that the answers received from server $i$ involve $W_{k,\tau}[1:L'/2]$, while the answers received from server $j$ involve $W_{k,\tau}[L'/2+1:L']$.

In the second stage, we consider all possible $2$-subsets of $\{E_{k,1},\ldots, E_{k,r}\}$. For each subset $\{E_{k,\tau_1}, E_{k,\tau_2}\}$, we apply $\Pi$ to message sums given by
\begin{align}
    \{W_{\ell,\tau_1}+W_{\ell,\tau_2}: (\ell,\tau_1)\in E_{k,\tau_1}, (\ell,\tau_2)\in E_{k,\tau_2}\}.
\end{align}
Out of these sums, $\binom{r-1}{1}$ contain $W_{k,1}$ as a term, and $\binom{r}{2}-(r-1) = \binom{r-1}{2}$ do not. For a fixed subset $\{E_{k,1}, E_{k,\tau}\}$, $\tau\in [2:r]$, we modify the queries of $\Pi$ such that, the answers from server $i$ involve $W_{k,1}[(\tau-1)L'+1:(\tau-1) L' +L'/2]+W_{k,\tau}[L'/2+1:L']$, while the answers from server $j$ involve $W_{k,1}[(\tau-1)L'+L'/2+1:\tau L']+W_{k,\tau}[1:L'/2]$. From here, we decode the $(r-1)L'$ desired message symbols, $W_{k,1}[L'+1:rL']$, by utilizing the side information from the first stage. Let the subsets that do not contain $E_{k,1}$ have a fixed ordering. Further, let $\{E_{k,\tau_1}, E_{k,\tau_2}\}$, $\tau_1\neq \tau_2, \tau_1, \tau_2 \in [2:r]$, be such that $E_{k,\tau_1}$ and $E_{k,\tau_2}$ occur at the $t_1$th and $t_2$th position ($t_1,t_2\in \left[\binom{r-1}{2}\right]$), respectively, among the ordered subsets. Then, we assume that the answers from server $i$ involve $W_{k,\tau_1}[(t_1-1)L'+1:(t_1-1)L'+L'/2]+W_{k,\tau_2}[(t_2-1)L'+1:(t_2-1)L'+L'/2]$, while the answers from server $j$ involve $W_{k,\tau_1}[(t_1-1)L'+L'/2+1:t_1L']+W_{k,\tau_2}[(t_2-1)L'+L'/2+1:t_2L']$. These sums of symbols act as side information for the next stage. 

For a general stage $3\leq\sigma\leq r-1$, we consider all possible $\sigma$-subsets of $\{E_{k,1},\ldots, E_{k,r}\}$. For a fixed subset $\{E_{k,\tau_1},\ldots, E_{k,\tau_{\sigma}}\}$, we apply $\Pi$ to message sums given by
\begin{align}
    \{W_{\ell,\tau_1}+\ldots +W_{\ell,\tau_{\sigma}}: (\ell,t)\in E_{k,t} ~\forall t\in \{\tau_1, \ldots, \tau_{\sigma}\}\},
\end{align}
to retrieve the symbol indices $\left[\sum_{s=1}^{\sigma-1} \binom{r-1}{s-1}L'+1:\sum_{s=1}^{\sigma-1}\binom{r-1}{s-1}L'+\binom{r-1}{\sigma-1}L'\right]$ of $W_{k,1}$. First, for the $\binom{r-1}{\sigma -1}$ subsets with $\tau_1=1$, and $\tau_2, \ldots, \tau_\sigma \in [2:r]$, let the respective subsets have a fixed ordering. Corresponding to the $t$th subset given by, $\{E_{k,1},E_{k,\tau_2}, \ldots, E_{k,\tau_\sigma}\}$, the answers recover the message symbols 
\begin{align}\label{sigma_message}
    W_{k,1}\left[\sum_{s=1}^{\sigma-1} \binom{r-1}{s-1}L'+(t-1)L'+1:\sum_{s=1}^{\sigma-1}\binom{r-1}{s-1}L'+tL'\right],
\end{align}
added to the $(\sigma-1)$-sum 
\begin{align}\label{eq:t_interference}
    \sum_{l=2}^\sigma W_{k,\tau_l}\left[(U_{\tau_l}-1)L'+1: U_{\tau_l}L\right],
\end{align}
where $[(U_{\tau_2}-1)L'+1:U_{\tau_2}L'], \ldots, [(U_{\tau_{\sigma}}-1)L'+1:U_{\tau_{\sigma}}L']$ in \eqref{eq:t_interference} are the symbols of the $(\sigma-1)$ message sums that were retrieved in stage $(\sigma-1)$, corresponding to the subset $\{E_{k,\tau_2}, \ldots, E_{k,\tau_\sigma}\}$, as side information. We follow the convention that the first $L'/2$ symbols of \eqref{sigma_message}, and the latter $L'/2$ symbols of \eqref{eq:t_interference} are contributed by server $i$, while the latter $L'/2$ symbols of \eqref{sigma_message}, and the first $L'/2$ symbols of \eqref{eq:t_interference} are contributed by server $j$. If $\tau_1\neq 1$, we fix the ordering of the $\binom{r-1}{\sigma}$ subsets. For a fixed subset $\{E_{k,\tau_1},\ldots, E_{k,\tau_r}\}$, let $t_l$ be the position of $\tau_l$ in this subset. Then, the answers recover the message symbols
\begin{align}
   \sum_{l=1}^\sigma W_{k,\tau_l}[(V_{\tau_l}-1)L'+1:V_{\tau_l}L'],
\end{align}
where $V_{\tau_l}=\sum_{s=1}^{\sigma-1}\binom{r-1}{s-1} + (t_l -1) $. In every such sum, the first $L'/2$ symbols are contributed by server $i$, while the last $L'/2$ symbols are contributed by server $j$.

In stage $r$, no new side information is created, and the remaining $L'$ symbols of $W_{k,1}$, i.e., $W_{k,1}\left[\sum_{s=1}^{r-1}\binom{r-1}{s-1}L'+1:2^{r-1}L'\right]$, are decoded by utilizing the side information from stage $r-1$. Thus, $W_{k,1}$ is decoded by downloading $D_{k,1} = \sum_{t=1}^r\binom{r}{t} D'$ message symbols. The same download cost is required for all $\tau\geq2$, yielding the rate
\begin{align}
    \frac{KL}{\sum_{\tau=1}^r \sum_{k=1}^{K'} D_{k,\tau}} &= \frac{KL'2^{r-1}}{rK'(2^r-1)D'}\\
    &= R^{\Pi}(G)(2-2^{1-r})^{-1}.
\end{align}
In every stage of the scheme, all possible sums of message subsets are involved, irrespective of the desired message index. When $\theta=k$, each invocation of the scheme $\Pi$ is locally private against the concerned servers $i$ and $j$. Finally, following the construction in \cite{our_journal2025}, the scheme takes care that the interference message symbols retrieved from server $i$ are queried in the subsequent stage from server $j$, summed with the desired message symbols, and vice versa. 

Next, we illustrate the scheme by extending the example on the bipartite graph $\mathbf{K}_{2,3}$ in Section~\ref{proof_complete_based} to $\mathbf{K}_{2,3}^{(2)}$ for $\theta=(1,1)$, by querying the servers 1 to 5. Let the $12$ symbols of each $W_{k,\tau}, ~k\in [6], ~\tau\in [2]$ be permuted independently and uniformly at random, and be labeled using the same letters as the previous example, without and with the prime symbol ($'$) when $\tau=1$, and $\tau=2$, respectively. From the scheme presented in Table~\ref{tab:answers_bipartite_multi}, the user retrieves $a_1, \ldots, a_{12}$, yielding the rate $\frac{12}{42}=\frac{2}{7}$.

\begin{table}[h]
    \centering
    \begin{tabular}{|c||c|c|c|c|c|}
    \hline
        stage & server 1 & server 2 & server 3 & server 4 & server 5\\
        \hline
        \multirow{4}{*}{$1$} & $a_1,b_1,c_1$& $d_3$ & $a_4,a_5,d_1,d_2$ & $b_2$& $c_2$\\
        & $a_2+b_2, a_3+c_2, b_3+c_3$ & & $a_6+d_3$ & & \\
         & $a'_1,b'_1,c'_1$& $d'_3$ & $a'_4,a'_5,d'_1,d'_2$ & $b'_2$& $c'_2$\\
        & $a'_2+b'_2, a_3'+c'_2, b'_3+c'_3$ & & $a'_6+d'_3$ & &\\
        \hline
        \multirow{4}{*}{$2$} & $a_7+a_4',b_7+b_7',c_7+c_7'$& $d_9+d'_9$ & $a_{10}+a'_1, a_{11}+a'_{2}$ & $b_8+b'_8$ & $c_8+c'_8$\\
        & $a_8+b_8+a'_5+b'_8$ & & $d_{7}+d'_{7},d_{8}+d'_{8}$ & &\\
        & $a_9+c_8+a'_6+c'_8$ &  & $a_{12}+d_9+a'_3+d'_9$ & &\\
        & $b_9+c_9+b'_9+c'_9$ & & & &\\
        \hline
    \end{tabular}
    \caption{Retrieval scheme for $\mathbf{K}_{2,3}^{(2)}$.}
    \label{tab:answers_bipartite_multi}
\end{table}

\subsubsection{Proof of Theorem~\ref{thm:gen_multigraph_ach}} 
The proof follows the ideas of Theorem~\ref{thm:gen_graph_ach}. Assume that the set of indices at server $n$ is specified as $\mathcal{I}_n =\bigcup_{\tau\in [r]} \{\ell_{1,\tau},\ell_{2,\tau},\ldots,\ell_{\deg(n),\tau}\}$, where $\ell_{1,\tau}<\ell_{2,\tau}<\ldots<\ell_{\deg(n),\tau}$ for all $\tau\in [r]$. Accordingly, let the messages at server $n$ be arranged in the form of the matrix
\begin{align}
    \bm{W}_n 
    &=
    \begin{bmatrix}
        W_{\ell_1,1} & W_{\ell_1,2} & \ldots & W_{\ell_1,r}\\
        W_{\ell_2,1} & W_{\ell_2,2} & \ldots & W_{\ell_2,r}\\
        \vdots & \vdots & \ddots& \vdots\\
        W_{\ell_{\deg(n)},1} & W_{\ell_{\deg(n)},2} & \ldots & W_{\ell_{\deg(n)},r}\\
    \end{bmatrix}.
\end{align}
The user generates $K$ bits $h_{s,\tau}$, $(s,\tau)\in [K']\times [r]$ independently, where each $h_{s,\tau}$ is picked uniformly at random from $\{0,1\}$. Let these bits be arranged into $n$ matrices as
\begin{align}
    \bm{H}_n & = 
    \begin{bmatrix}
    h_{\ell_1,1} & h_{\ell_1,2} & \ldots & h_{\ell_1,r}\\
        h_{\ell_2,1} & h_{\ell_2,2} & \ldots & h_{\ell_2,r}\\
        \vdots & \vdots & \ddots & \vdots\\
        h_{\ell_{\deg(n)},1} & h_{\ell_{\deg(n)},2} & \ldots & h_{\ell_{\deg(n)},r}
    \end{bmatrix}.
\end{align}Let $\theta = (k,1)$ with $W_{k,1}, \ldots, W_{k,r}$ stored at servers $i$ and $j$. Further, let $[W_{k,1}, \ldots, W_{k,r}]$ be the $p$th row of $\bm{W}_i$, and the $m$th row of server $j$. Like the simple graph scheme, only servers $i$ and $j$ ($i<j$) receive a query given by 
\begin{align}
    Q_i^{[k,1]} = \bm{H}_i, \qquad Q_j^{[k,1]} = \bm{H}_j + \bm{e}_m\bm{e}_1^\top,
\end{align}
where $\bm{H}_i(p,:)=\bm{H}_j(m,:)=\begin{bmatrix}h_{k,1} & h_{k,2} & \ldots & h_{k,r}\end{bmatrix}$, and $\bm{e}_m\bm{e}_1^\top$ is a $\deg(j)\times r$ matrix of all zeros with a $1$ at the $(m,1)$th coordinate. The answer $A_n^{[k,1]}$ returned by  servers $i$ and $j$ is a vector whose $\nu$th entry $A_n^{[k,1]}(\nu)$ is given by
\begin{align}
    A_n^{[k,1]}(\nu) &= \begin{cases}
       Q_n^{[k,1]}(\nu,:)\bm{W}_n(\nu,:)^\top, &Q_n^{[k,1]}(\nu,:)\neq\boldsymbol{0}^\top ,\\
       \emptyset, &  Q_n^{[k,1]}(\nu,:)=\boldsymbol{0}^\top,
    \end{cases}
\end{align}
where $\boldsymbol{0}$ is the $r$-length column vector of all zeroes. The corresponding download cost $D_{k,1}$ is given by the total length of answers $A_i^{[k,1]}$ and $A_j^{[k,1]}$, averaged over all realizations of $\{h_{\ell,\tau}, ~ \ell\in \mathcal{I}_i \cup \mathcal {I}_j,~ \tau\in [r]\}\in \{0,1\}^{r(\deg(i)+\deg(j))}$.  

First, we compute the average answer length for the queries generated from $h_{k,1}, \ldots, h_{k,r}$, which correspond to $Q_i^{[k,1]}(p,:)$ and $Q_j^{[k,1]}(m,:)$, for server $i$ and server $j$, respectively. A single bit, i.e., $W_{k,1}$ is returned either by server $j$ alone, if the $p$th query row, $Q_i^{[k,1]}(p,:)=\bm{H}_i(p,:)=\boldsymbol{0}^\top$  or by server $i$ alone, if the $m$th query row, $Q_j^{[k,1]}(m,:)=\bm{H}_j(m,:)+\bm{e}_1^\top=\boldsymbol{0}^\top$. Notice that each of these events occurs with probability $2^{-r}$. Otherwise, each of the servers $i$ and $j$ returns a single bit, i.e., $2$ bits in total, whenever both the query vectors $Q_i^{[k,1]}(p,:)$ and $Q_j^{[k,1]}(m,:)$ are non-zero. This occurs with probability $1-2^{1-r}$. Next, for the queries generated from $\{h_{\ell,\tau}, \ell\in \mathcal{I}_i \cup\mathcal{I}_j\setminus \{k\}, \tau \in[r]\}$, let $X_i\in [0:\deg(i)-1]$ and $X_j\in [0,\deg(j)-1]$ be the number of rows in $\bm{H}_i$ and $\bm{H}_j$, respectively, that are not $\boldsymbol{0}^\top$, excluding the row of $[h_{k,1},\ldots,h_{k,r}]$. Picking any such row from $\bm{H}_i$ and $\bm{H}_j$ (say, the $\nu$th row), we have
\begin{align}
    \mathbb{P}\left(\bm{H}_i(\nu,:) \neq \boldsymbol{0}^\top\right) &= 1- \mathbb{P}\left(\bm{H}_i(\nu,:) = \boldsymbol{0}^\top\right)=1-2^{-r}, \\
      \mathbb{P}\left(\bm{H}_j(\nu,:) \neq \boldsymbol{0}^\top\right) &= 1- \mathbb{P}\left(\bm{H}_j(\nu,:) = \boldsymbol{0}^\top\right)=1-2^{-r}.
\end{align}
Since the random variables are independent, the entries of the rows are mutually independent. This yields
\begin{align}
    X_i &= \sum_{\nu\in [\deg(i)]\setminus \{p\}}\mathds{1}\left(\bm{H}_i(\nu,:) \neq \boldsymbol{0}^\top\right), \\   
    X_j &= \sum_{\nu\in [\deg(j)]\setminus\{m\}}\mathds{1}\left(\bm{H}_j(\nu,:) \neq \boldsymbol{0}^\top\right).
\end{align}
Thus, we have
\begin{align}
    D_{k,1} &= 1\cdot 2^{1-r}+2\cdot(1-2^{1-r})+\mathbb{E}\left[X_i\right]+ \mathbb{E}\left[X_j\right]\\
    &=2-2^{1-r}+(\deg(i)-1)(1-2^{-r})+(\deg(j)-1)(1-2^{-r})\\
    &= (1-2^{-r})\left(\deg(i)+\deg(j)\right). \label{eq:download_per_index}
\end{align}
The expression in \eqref{eq:download_per_index} for $D_{k,\tau}$ holds for all $\tau \in [r]$. Summing over $k$ and $\tau$ yields the total download as
\begin{align}
        \sum_{k=1}^{K'}\sum_{\tau=1}^r D_{k,\tau} &= \sum_{k=1}^{K'}r(1-2^{-r})(\deg(i)+\deg(j))\\
        &=r(1-2^{-r}) \sum_{\{i,j\}\in E}(\deg(i)+\deg(j))\\
        &= r(1-2^{-r})\sum_{n\in [N]} \deg(n)^2 \label{total_download_gen_multigraph},
\end{align}
which gives the desired result.

For Theorems \ref{thm:multicycle_bounds} and \ref{thm:multipath_bounds}, the lower bounds follow from Theorem~\ref{thm:srp_multigraph_ach}. For the upper bounds, we need the following lemma, which is a modified form of \cite[Lemma  23]{our_journal2025}, relaxed for local privacy, with respect to the indices $\theta$ for which this constraint needs to be satisfied.

\begin{lemma}[Lemma 23, \cite{our_journal2025}]\label{lem:multigraph_decodability_lemma}
    For any two servers $i,j$ with $\{i,j\}\in E$, and for any $\theta\in \mathcal{I}_i \cap \mathcal{I}_j$, a local PIR scheme on the $r$-multigraph $G^{(r)}$,  satisfies
    \begin{align}
        H(A_i^{[\theta]}|\mathcal{W}\setminus \mathcal{W}_{i,j}, \mathcal{Q})+H(A_j^{[\theta]}|\mathcal{W}\setminus \mathcal{W}_{i,j}, \mathcal{Q})\geq L\left(2-2^{1-r}\right).
    \end{align}
\end{lemma}

To prove the upper bounds, we find a lower bound on the download cost for each desired index, and deviate slightly from the simple graph case as follows. Starting with a fixed index, $\theta = (s,\tau)$, where $s\in [N], ~\tau\in [r]$, we obtain
\begin{align}
    D_{s,\tau}&\geq L+ I(\mathcal{W}\setminus \{W_{s,\tau}\};A_{[N]}^{[s,\tau]}|W_{s,\tau},\mathcal{Q})\\
    &= L+ I(\mathcal{W}\setminus \mathcal{W}_{s,s+1}, \mathcal{W}_{s,s+1}\setminus\{W_{s,\tau}\};A_{[N]}^{[s,\tau]}|W_{s,\tau},\mathcal{Q})\\
    &=  L+ I( \mathcal{W}_{s,s+1}\setminus\{W_{s,\tau}\};A_{[N]}^{[s,\tau]}|W_{s,\tau},\mathcal{Q}) +  I( \mathcal{W}\setminus\mathcal{W}_{s,s+1};A_{[N]}^{[s,\tau]}|\mathcal{W}_{s,s+1},\mathcal{Q}) \label{eq:set this to zero}\\
    &\geq L + I( \mathcal{W}\setminus\mathcal{W}_{s,s+1};A_{[N]}^{[s,\tau]}|\mathcal{W}_{s,s+1},\mathcal{Q}),\label{starting_point_multigraph}
\end{align}
where \eqref{starting_point_multigraph} holds by setting the first term in \eqref{eq:set this to zero} to zero. For $r>1$, this dropped term contributes to the gap between the lower and upper bounds. The proofs of the latter (Theorems~\ref{thm:multicycle_bounds} and \ref{thm:multipath_bounds}) appear next.

\subsubsection{Proof of Theorem~\ref{thm:multicycle_bounds}}
The storage at server $n\in [N]$ is given by the $2r$ messages,
\begin{align}
    \mathcal{W}_n &=\mathcal{W}_{n-1,n}\cup \mathcal{W}_{n,n+1}\\
    &= \{W_{n-1,\tau}, W_{n,\tau},~ \tau\in [r]\},
\end{align} 
where the indices are added and subtracted modulo $N$. The second term in \eqref{starting_point_multigraph} can be lower-bounded as
\begin{align}
    I(& \mathcal{W}\setminus\mathcal{W}_{s,s+1};A_{[N]}^{[s,\tau]}|\mathcal{W}_{s,s+1},\mathcal{Q})\notag\\
    &\geq I(\mathcal{W}_{s-1,s};A_s^{[s-1,\tau]}|\mathcal{W}_{s,s+1},\mathcal{Q})+ I(\mathcal{W}_{s+1,s+2};A_{s+1}^{[s+1,\tau]}|\mathcal{W}_{s,s+1},\mathcal{W}_{s-1,s},\mathcal{Q})\\
    &= H(A_s^{[s-1,\tau]}|\mathcal{W}_{s,s+1},\mathcal{Q})+ H(A_{s+1}^{[s+1,\tau]}|\mathcal{W}_{s,s+1},\mathcal{W}_{s-1,s},\mathcal{Q})\\
    &\geq H(A_s^{[s-1,\tau]}|\mathcal{W}\setminus \mathcal{W}_{s-1,s},\mathcal{Q})+ H(A_{s+1}^{[s+1,\tau]}|\mathcal{W}\setminus \mathcal{W}_{s,s+1},\mathcal{Q}).
\end{align}
Now, summing over all $s,\tau$, following the steps in Section~\ref{proof:cyclic_converse} and using Lemmas~\ref{lem:user_priv}, \ref{lem:answer_indep_randomness_given_query} and \ref{lem:multigraph_decodability_lemma}, we obtain
\begin{align}
    \sum_{s=1}^N \sum_{\tau=1}^r D_{s,\tau} &\geq NrL+Nr(2-2^{1-r})L\\
    &= NrL(3-2^{1-r}).
\end{align}

\subsubsection{Proof of Theorem~\ref{thm:multipath_bounds}} 
Recall that the storage of $\mathbf{P}_N^{(r)}$ is 
\begin{align}
    \mathcal{W}_n&=
    \begin{cases}
    \mathcal{W}_1 = \{W_{1,\tau}, ~\tau\in [r]\}, & n=1,\\
    \mathcal{W}_{n-1,n}\cup\mathcal{W}_{n,n+1}=\{W_{n-1,\tau}, W_{n,\tau},~ \tau\in [r]\}, & n\in [2:N-2],\\
    \mathcal{W}_{N-1}=\{W_{N-1,\tau}, ~\tau\in [r]\}, & n=N-1.
    \end{cases}    
\end{align}Starting from \eqref{starting_point_multigraph}, and following the arguments in Section~\ref{proof:path_converse}, we lower bound the interference for all $\tau\in [r]$ as
\begin{align}
    I(& \mathcal{W}\setminus\mathcal{W}_{s,s+1};A_{[N]}^{[s,\tau]}|\mathcal{W}_{s,s+1},\mathcal{Q})\notag\\&\geq
    \begin{cases}
    H(A_{s+1}^{[s+1,\tau]}|\mathcal{W}\setminus \mathcal{W}_{s+1,s+2}, \mathcal{Q}), & s=1,\\
    H(A_s^{[s-1,\tau]}|\mathcal{W}\setminus \mathcal{W}_{s-1,s},\mathcal{Q}) + H(A_{s+1}^{[s+1, \tau]}|\mathcal{W}\setminus \mathcal{W}_{s+1,s+2},\mathcal{Q}),& s \in [2:N-2],\\
    H(A_{s}^{[s-1, \tau]}|\mathcal{W}\setminus \mathcal{W}_{s-1,s}, \mathcal{Q}), & s=N-1.
    \end{cases}
\end{align}
Summing over all $(s,\tau)\in [N-1]\times [r]$,
\begin{align}
    \sum_{s=1}^{N-1}\sum_{\tau=1}^r D_{s,\tau}&\geq (N-1)rL + (N-3)rL(2-2^{1-r})\\ 
    &= rL\left(N-3)(3-2^{1-r} + 2 \right).
\end{align}

\section{Conclusion}\label{sec:conclude}
In this paper, we studied graph-based private information retrieval from the perspective of local privacy for simple graphs and multigraphs. On one hand, fully-replicating the database at all servers improves the capacity with the increased number of servers. On the other hand, with the massive amount of data, realizing full-replication becomes more and more expensive and essentially untenable. Therefore, settling for a compromise of (hyper)-graph replicated storage for PIR is a judicious model to avoid extremely large replication costs. However, the original definition of PIR under this setup is too conservative and causes degradation in capacity as more servers are included in the system. This is because inclusion of servers inevitably adds more messages to the PIR system, further amplifying the privacy requirement. To circumvent this issue, our local privacy formulation offers a reasonable privacy definition, where the retrieval of a message is hidden from a server if and only if that message is stored in that sever. 

The relaxation of privacy yields a natural improvement over the communication efficiency of PIR, and the results are more encouraging than those of the original treatment of graph-replicated PIR. Moreover, as $N$ increases, the local PIR capacity decays at a slower rate than the standard PIR capacity. For simple graphs, we proposed capacity lower bounds that are at least $O(\sqrt{N})$ times superior to the rates of standard PIR, whereas cyclic and path graphs exhibit $O(N)$ times improvement in the capacity. For multigraphs, the rates achieved through our proposed schemes are the rates of the respective simple graphs, multiplied by the factor $(2-2^{1-r})^{-1}$. Further work is needed to characterize the exact/approximate capacity of graphs other than star, cyclic and path, to understand the limits of rate improvement of local PIR against PIR. Similarly, in the domain of multigraphs, characterizing the capacity bounds of most multigraphs remains open.

\bibliographystyle{unsrt}
\bibliography{references.bib}
\end{document}